\pdfoutput=1
\documentclass[aps,prd,floatfix,amsmath,amssymb,onecolumn,groupedaddress,11pt]{revtex4}
\usepackage{graphicx}
\usepackage{amssymb}
\usepackage{epstopdf}
\usepackage{subfigure}
\usepackage{array}
\usepackage{color}
\usepackage{pdfsync}
\usepackage{amsmath}
\usepackage{hyperref}
\newcommand{\Tr}[0]{\text{Tr}}
\newcommand{\STr}[0]{\text{STr}}

\newcommand{\1}[0]{\textbf 1}

\newcommand{\st}[0]{\mathcal{S}_{3}}

\newcommand{\Hu}[1]{\mathcal H_{u#1}}
\newcommand{\Hd}[1]{\mathcal H_{d#1}}

\begin{document} 

\title{Flavored Gauge Mediation with Discrete Non-Abelian Symmetries}
\author{Lisa L.~Everett}
\email{leverett@wisc.edu} 
\author{Todd S.~Garon}
\email{tgaron@wisc.edu}  
\affiliation{Department of Physics, University of Wisconsin-Madison, Madison, WI 53706}
\date{\today}

\begin{abstract}
We explore the model-building and phenomenology of flavored gauge mediation models of supersymmetry breaking in which the electroweak Higgs doublets and the $SU(2)$ messenger doublets are connected by a discrete non-Abelian symmetry.  The embedding of the Higgs and messenger fields into representations of this non-Abelian Higgs-messenger symmetry results in specific relations between the Standard Model Yukawa couplings and the messenger-matter Yukawa interactions.  Taking the concrete example of an $\mathcal{S}_3$ Higgs-messenger symmetry, we demonstrate that while the minimal implementation of this scenario suffers from a severe $\mu/B_\mu$ problem that is well-known from ordinary gauge mediation,  expanding the Higgs-messenger field content allows for the possibility that $\mu$ and $B_\mu$ can be separately tuned, allowing for the possibility of phenomenologically viable models of the soft supersymmetry breaking terms.  We construct toy examples of this type that are consistent with the observed 125 GeV Higgs boson mass.  

\end{abstract}
\maketitle 

\section{Introduction}

The theoretical paradigm of TeV-scale supersymmetry (SUSY)  continues to be one of the best-motivated candidates for new physics that can be probed extensively at the Large Hadron Collider (LHC) (see e.g.~\cite{Martin:1997ns,Chung:2003fi} for reviews).  Indeed, supersymmetric extensions of the Standard Model (SM) such as the minimal supersymmetric standard model (MSSM) have already been subjected to stringent tests at the LHC, both from direct constraints such as the non-observation of superpartners to date, with limits on colored superpartners that reach well into the TeV region, and constraints from the 2012 discovery \cite{Aad:2012tfa,Chatrchyan:2012xdj} of a new scalar particle that is compatible with the SM Higgs boson.  While the Higgs mass $m_h\approx 125$ GeV is within the allowed range of (perturbative) supersymmetric models, its relatively high value requires either (i) large stop  mixing or very heavy stops in the MSSM (see e.g.~\cite{Carena:2002es}) or (ii) extended Higgs sectors.   Together, the data has placed severe limits on the viable regions of the vast (more than 100-dimensional) parameter space of the MSSM, largely ruling out many minimal scenarios for the soft supersymmetry breaking parameters.

Of the possible directions to explore in SUSY model-building in light of the experimental bounds, the gauge-mediated supersymmetry breaking framework \cite{gauge1,gauge2,gauge3,Giudice:1998bp} is particularly compelling due to its lack of sensitivity to unknown UV physics as compared to the gravity mediation framework.  
 In the context of the MSSM, however,  the minimal implementation of this idea, known as minimal gauge mediation, is particularly constrained by the LHC Higgs measurements \cite{Draper:2011aa,Arbey:2011ab,Ajaib:2012vc}. The reason is that minimal gauge mediation models predict family-universal scalar masses and vanishing soft trilinear scalar parameters ($A$ terms) at the messenger scale, requiring a significant amount of renormalization group (RG) evolution to generate the needed stop mixing.  Even when this structure can be obtained, in this class of models it is generally the case that the Higgs mass bound requires a very heavy superpartner spectrum  that is largely inaccessible at the LHC. 

Therefore, it is desirable to go beyond minimal gauge mediation and consider more intricate models in which the messengers have nontrivial Yukawa couplings to the SM fields, as first discussed in  \cite{gauge3,Giudice:1998bp,Chacko:2001km} and more recently considered in  \cite{Shadmi:2011hs,Evans:2011bea,Kang:2012ra,Craig:2012xp,Albaid:2012qk,Abdullah:2012tq,Perez:2012mj,Byakti:2013ti,Evans:2013kxa,Calibbi:2013mka,Galon:2013jba,Fischler:2013tva,Calibbi:2014yha,Ierushalmi:2016axs}.  
With a standard messenger sector of some number of vectorlike pairs of $(\mathbf{5}, \bar{\mathbf{5}})$ of $SU(5)$ (as typically assumed so as not to spoil the approximate gauge coupling unification of the MSSM), there are many possible renormalizable couplings of the messengers to the matter fields, as enumerated e.g.~in \cite{Evans:2013kxa,Jelinski:2015voa}.  The generic outcome of the presence of one or more of such couplings is that the soft terms now include two-loop contributions to the scalar masses and one-loop contributions to the $A$ terms that depend on the messenger Yukawa couplings (and depending on the model, there can also be one-loop contributions to the scalar masses).  Hence, while the flavor-blind structure of minimal gauge mediation is sacrificed, what has been gained is a much greater ease in accommodating the Higgs mass constraints, and thus in constructing viable models of the MSSM parameter space.  

One class of interesting models of this type is one in which the $SU(2)$ messenger doublets of the $(\mathbf{5}, \bar{\mathbf{5}})$ pairs couple to the SM fields in a similar way to the electroweak Higgs fields $H_{u,d}$.  In such models, which are known as ``flavored gauge mediation" models (see e.g.~\cite{Shadmi:2011hs,Abdullah:2012tq,Calibbi:2013mka,Galon:2013jba,Fischler:2013tva,Calibbi:2014yha,Ierushalmi:2016axs}), the underlying mechanism for generating the SM Yukawa couplings, such as via a horizontal $U(1)$ symmetry, should also play a dominant role in the structure of the messenger Yukawa couplings.  One immediate situation then is to address the impact of the flavor-violating contributions to the soft supersymmetry breaking parameters (see e.g.~\cite{Gabbiani:1988rb,Hagelin:1992tc,Gabbiani:1996hi,Raz:2002zx}) that are a generic consequence of this model structure. Models of this type can be  constructed such that new contributions are consistent with minimal flavor violation (MFV), and thus are safe from large flavor-violating effects beyond the SM.  In cases in which there is not precise alignment of this type, viable models can also be constructed in which the messenger Yukawas can share the same parametric suppression as the SM Yukawas, which can also result in acceptably small flavor violation \cite{Shadmi:2011hs,Abdullah:2012tq,Calibbi:2013mka,Galon:2013jba}.   While phenomenologically acceptable constructions are not guaranteed,  it has been shown that the potential flavor and CP violation in this class of models is more strongly suppressed in general than indicated by naive estimates due to the special structure of the soft terms as governed by a (softly) broken ``messenger number" symmetry  \cite{Calibbi:2014yha,Ierushalmi:2016axs}.

Within the flavored gauge mediation framework, one novel model-building direction is to consider scenarios in which the $SU(2)$ messenger doublets and the electroweak Higgs doublets transform as multiplets under a discrete non-Abelian symmetry.  The idea that this discrete non-Abelian symmetry is also the horizontal family symmetry that governs the SM and messenger Yukawa couplings was first proposed by proposed by Perez, Ramond, and Zhang \cite{Perez:2012mj} (hereafter referred to as ``PRZ").  In the PRZ approach, the field that breaks the family symmetry also breaks supersymmetry, resulting in soft supersymmetry breaking terms with a nontrivial flavor dependence that originates both from the details of the family symmetry breaking and the mixing of the Higgs and messenger fields.   After enumerating the constraints for generating reasonable soft supersymmetry breaking terms, PRZ constructed a toy two-generation model in which the non-Abelian symmetry group is the $\mathcal{S}_3$ symmetry group.  This model achieves hierarchical SM Yukawas and predicts that the messenger Yukawas obey an inverted hierarchy; the resulting off-diagonal flavor-violating couplings of the soft supersymmetry breaking terms can be reduced by RG effects from the messenger scale to the electroweak scale.
PRZ's approach is striking in that it dispels the standard folklore that the unification of family symmetry breaking and supersymmetry breaking inevitably leads to excessive flavor violation.  Their approach also suggests new flavor model-building directions in which the Higgs fields transform nontrivially under the family symmetry group.

In this paper, we continue the exploration of the idea that the electroweak and messenger doublets are connected via a discrete non-Abelian symmetry, which we also take to be  $\mathcal{S}_3$ for simplicity, and explore possibilities for constructing potentially viable models with three generations.   Our study deviates from the PRZ framework in that we consider two possibilities for the $\mathcal{S}_3$ symmetry: (i) it is just a symmetry relating the Higgs and messenger doublets, and hence the SM fields are  $\mathcal{S}_3$ singlets, and (ii) it is (part of) the full family symmetry that governs the SM Yukawa couplings, and hence the SM fields are embedded in nontrivial $\mathcal{S}_3$ representations.  Clearly, the choice made will dramatically affect the possibilities for the SM and messenger Yukawa couplings, and thus the structure of the resulting soft terms.    Our focus will not be on constructing fully viable three-generation models of the SM and messenger Yukawas, but instead on the structure of the soft terms in each case and the resulting constraints on the superpartner mass spectrum.  Hence, we will focus on third generation couplings, and defer a more complete study for future work.
 
We will see immediately that in this approach, we will generally be confronted by a severe $\mu/B_\mu$ problem that must be addressed to have any hope of obtaining a realistic theory.  The presence of a $\mu/B_\mu$ problem is a well-known problem in gauge mediation \cite{Dvali:1996cu,Giudice:2007ca} (see also \cite{Giudice:1998bp} for an overview).  Here it is more severe than usual because of the necessity of coupling the Higgs fields to the supersymmetry breaking field at the renormalizable level, since the Higgs and doublet messengers are connected by the discrete non-Abelian symmetry.  We will show that while a minimal implementation of this framework leads inevitably to this severe $\mu/B_\mu$ problem, viable scenarios can be constructed when the messenger sector is enlarged to include different representations of the Higgs-messenger fields with respect to the $\mathcal{S}_3$ symmetry, such that the $\mu$ and $B_\mu$ parameters can be {\it separately} tuned to acceptable values.  This therefore is not a compelling solution to the $\mu/B_\mu$ problem of gauge mediation, but it does at least allow for the possibility of viable (albeit tuned) models of the soft supersymmetry breaking mass parameters of the MSSM.

The structure of the paper is as follows.  We begin in the next section with an overview of our theoretical model-building framework, and discuss the ways in which it has similarities and differences to the PRZ approach.  We then demonstrate how a minimal implementation of the Higgs-messenger sector results in the severe $\mu/B_\mu$ problem just mentioned, and discuss possible resolutions of the issue.  In the following section, we next present an example of an enlarged Higgs-messenger field content that allows for separate adjustments of $\mu$ and $B_\mu$. We follow this discussion with examples of different assignments of the SM fields into representations of the $\mathcal{S}_3$ symmetry, and discuss the resulting impact on the Yukawa couplings of the SM fields to the electroweak Higgs fields and the messenger fields.  For each example, we investigate the phenomenology of the resulting soft supersymmetry breaking terms, and investigate patterns of superpartner mass spectra that are consistent with the 125 GeV Higgs mass.  Finally, we summarize and discuss prospects future model-building directions along these lines.

\section{Theoretical Framework}
\label{backgroundsection}

\subsection{General considerations}

Our framework for exploring flavored gauge mediation is as follows.  We assume the presence of a discrete non-Abelian symmetry that we denote by $\mathcal{G}$.  The key features of $\mathcal{G}$ are as follows.  First, 
$\mathcal{G}$ relates the chiral superfields that will later be identified as the electroweak Higgs fields $H_{u,d}$ and the $SU(2)$ doublet messengers $M_{ui,di}$, in which $i=1,2,\ldots, N$, into a set of representations of $\mathcal{G}$, that we denote collectively as follows:
\begin{eqnarray}
\mathcal{H}_{u}= \left (\begin{array}{c} \mathcal{H}_{u1}\\ \mathcal{H}_{u2}\\ \vdots \\ \mathcal{H}_{u N+1}\end{array} \right ) = \mathcal{R}_u \left (\begin{array}{c} H_u\\M_{u1} \\ \vdots \\ M_{uN} \end{array} \right ), \qquad  \mathcal{H}_{d}= \left (\begin{array}{c} \mathcal{H}_{d1}\\ \mathcal{H}_{d2}\\ \vdots \\ H_{dN+1} \end{array} \right ) = \mathcal{R}_d \left (\begin{array}{c} H_d\\M_{d1}\\ \vdots  \\ M_{dN} \end{array} \right ),
\label{higgs_s3}
\end{eqnarray}
in which the $\mathcal{R}_{u/d}$ are rotation matrices that are obtained upon diagonalizing the mass matrices of the Higgs/doublet messenger sector of the theory.
In the above, we note that $\mathcal{H}_{u,d}$ can either represent a single set of multiplets of $\mathcal{G}$, or it can represent a collection of them, as we will later explore in greater detail.   We will take the $SU(3)$ triplet messengers $T_{ui,di}$, which are needed to generate gluino masses, to be $\mathcal{G}$ singlets for simplicity.  Recall that the messenger triplets $T_{ui,di}$ and messenger doublets $M_{ui,di}$ together form $N$ vectorlike pairs of $\mathbf{5}$, $\overline{\mathbf{5}}$ representations of $SU(5)$.

Second, the spontaneous breaking of $\mathcal{G}$ is due at least partially to a field or fields that break supersymmetry in the hidden sector.  In other words, in the supersymmetry breaking sector, which consists of superfields that develop both  scalar and F-component vacuum expectation values (vevs), there is at least one field that has a nontrivial $\mathcal{G}$ representation.  We will also assume that there is a supersymmetry-breaking field $X_T$ that is a singlet with respect to $\mathcal{G}$;  we assume throughout that this field couples only to the ($SU(3)_c$ triplet and $\mathcal{G}$-singlet) messengers $T_{ui,di}$ as follows:
\begin{eqnarray}
W_T= \lambda_T X_T T_{ui} T_{di},
\end{eqnarray} 
such that when $X_T$ develops vacuum expectation values in the scalar and F-components
\begin{eqnarray}
\langle \lambda_T X_T \rangle = M _T+\theta^2 F_T.
\end{eqnarray}
We will later specify options for the coupling of $\mathcal{H}_{u,d}$ to the supersymmetry breaking sector.
In addition, depending on the model in question, there may be other ``flavon" supermultiplets that are SM singlets charged under $\mathcal{G}$ that acquire scalar vevs, but do not participate in supersymmetry breaking.  These fields may be needed to generate nontrivial Yukawa couplings to the matter fields, as discussed below.
 
The role of $\mathcal{G}$ for the observable sector fields other than the electroweak Higgs doublets has not yet been specified. In this sector, which we presume for simplicity consists solely of the MSSM matter and gauge supermultiplets, the question of whether the matter fields are also charged with respect to $\mathcal{G}$ or not will be a model-building choice that will have a significant impact on the resulting MSSM and messenger Yukawa couplings.  We will explore different options in this work.  

The diagonalization of the messenger-Higgs sector and the identification of the electroweak Higgs fields $H_{u,d}$ and the (heavier) messenger doublets $M_{ui,di}$, together with the mixing matrices $\mathcal{R}_{u,d}$, will play a critical role in the structure of the MSSM and messenger Yukawa couplings.  In a schematic form, the superpotential couplings involving the MSSM matter superfields (in self-evident notation) and $\mathcal{H}_{u,d}$ of Eq.~(\ref{higgs_s3}) take the form (neglecting neutrino masses for simplicity):
\begin{eqnarray}
W_Y=(y_u Q\overline{u} \mathcal{H}_u)+
 (y_d Q\overline{d} \mathcal{H}_d)+(y_e L\overline{e} \mathcal{H}_d),
\label{yukawas}
\end{eqnarray}
in which family indices have been suppressed for simplicity, and the parentheses denote contractions of $\mathcal{G}$.  $y_{u,d,e}$ represent prefactors that may either be Yukawa couplings of renormalizable superpotential couplings or effective couplings originating from higher-dimensional operators.  Upon supersymmetry breaking and the breakdown of $\mathcal{G}$, the effective Yukawa couplings of the MSSM fields to the MSSM Higgs fields $H_{u,d}$ and to the heavy messenger doublets $M_{u,di }$, take the form
\begin{eqnarray}
W_Y = Y_u Q\overline{u} H_u+Y_d Q\overline{d} H_d+ Y_e L\overline{e} H_d+
Y_{ui}^\prime Q\overline{u} M_{ui}+Y_{di}^\prime Q\overline{d} M_{di}+ Y_{ei}^\prime L_m\overline{e} M_{di}.
\label{effyukawas}
\end{eqnarray}
While the details of the relations between the MSSM Yukawa couplings $Y_u$, $Y_d$, $Y_e$ and the messenger Yukawa couplings $Y^\prime_{ui}$, $Y^\prime_{di}$, and $Y^\prime_{ei}$ will depend in detail on the model, the two sets of couplings are generally related and depend on various entries of the unitary matrices $\mathcal{R}_{u,d}$.   As is well known, messenger Yukawa couplings of this type result in additional contributions to the soft terms beyond those present in minimal gauge mediation.  These corrections have been computed for example in \cite{Abdullah:2012tq} and are given for completeness in full three-family structure in Appendix \ref{flavorcorrections}.

The messenger Yukawas are not necessarily diagonal in the same basis as the MSSM Yukawas, such that the constraints from experimental bounds on flavor-changing processes must be considered carefully in each case.  Analyses of the constraints from flavor physics on certain classes of flavored gauge mediation scenarios have been presented in \cite{Calibbi:2013mka}.   A complete analysis of the question of the viability of full three-family models necessarily involves the complete modeling of the SM Yukawa couplings, which we do not pursue here.  Instead, we consider toy scenarios, depending on ways in which the MSSM matter fields are embedded within representations of the non-Abelian discrete symmetry group $\mathcal{G}$, and focus on the effects on the MSSM soft terms.

In this paper, we presume that the couplings between the messengers and the MSSM fields as given in Eq.~(\ref{effyukawas}) are the {\it only} direct interactions between the sectors (see \cite{Evans:2013kxa} for a classification of additional terms that can be in principle be present). This will generically require the presence of symmetries in addition to $\mathcal{G}$  to ensure that such additional messenger-matter couplings are absent; we leave the possibility of including them to future work.

Up to this point, we have left $\mathcal{G}$ unspecified.  We will now consider the concrete case that $\mathcal{G}=\mathcal{S}_3$, the permutation group on three objects. The group theory of $\mathcal{S}_3$ can be found in many references (see e.g.~\cite{Perez:2012mj}); here we just mention a few main features for completeness.  $\mathcal{S}_3$ contains three irreducible representations, the singlet $\mathbf{1}$, a one-dimensional representation $\mathbf{1}^\prime$, and a doublet, $\mathbf{2}$, with tensor products 
\begin{eqnarray}
\mathbf{1}\otimes \mathbf{2}=\mathbf{2}, \qquad \mathbf{1}^\prime \otimes  \mathbf{2}=\mathbf{2}, \qquad  \mathbf{2}\otimes \mathbf{2}=\mathbf{1}\oplus \mathbf{1}^\prime\oplus\mathbf{2}.
\end{eqnarray}
We will use the same presentation as PRZ \cite{Perez:2012mj}.  In this basis, the singlet representation obtained from the tensor products of either two doublets or three doublets is as follows:
\begin{eqnarray}
(\mathbf{2} \otimes \mathbf{2})_\mathbf{1}&=& \left [\left (\begin{array}{c} a_1 \\ a_2 \end{array} \right ) \otimes  \left (\begin{array}{c} b_1 \\ b_2 \end{array} \right ) \right ]_\mathbf{1} = a_1 b_2 +a_2b_1. \\
(\mathbf{2} \otimes \mathbf{2}\otimes \mathbf{2})_\mathbf{1}&=& \left [\left (\begin{array}{c} a_1 \\ a_2 \end{array} \right ) \otimes  \left (\begin{array}{c} b_1 \\ b_2 \end{array} \right )\otimes  \left (\begin{array}{c} c_1 \\ c_2 \end{array} \right ) \right ]_\mathbf{1} =a_1b_1c_1+a_2b_2 c_2.
\label{cg}
\end{eqnarray}
Here we will restrict ourselves for simplicity to the case in which the fields in our model framework are either the $\mathbf{1}$ or $\mathbf{2}$ representations of $\mathcal{S}_3$, in which case Eq.~(\ref{cg}) provide us with the relations needed to construct $\mathcal{S}_3$ invariants.

\subsection{A minimal Higgs-messenger sector and the $\mu/B_\mu$ problem}
We now turn to the model-building of the Higgs-messenger sector. We will first consider a minimal implementation of this sector, in which by minimal we mean the number of degrees of freedom; this scenario was also considered in PRZ \cite{Perez:2012mj}.  The first ingredient is the introduction of a hidden sector field $X_H$, which is taken to be a $\mathbf{2}$ of $\mathcal{S}_3$.  The next ingredients are the assignment of Higgs-messenger fields $\mathcal{H}_u$ and $\mathcal{H}_d$ to doublet representations of $\mathcal{S}_3$ as well.  These charge assignments are shown in Fig.~\ref{tab:10} (here we neglect to show $X_H$ and the triplets $T_{ui,di}$).
\begin{table}[htbp]
   \centering
    \begin{tabular}{c|cc|c}
     & $\Hu{\mathbf 2}$ & $\Hd{\mathbf 2}$ &$X_H$\\
    \hline
    $\mathcal{S}_3$ &$\mathbf 2 $&  $\mathbf 2 $& \textbf 2\\
   \end{tabular}
   \caption{$\mathcal{S}_3$ charge assignments for a minimal Higgs-messenger sector as studied in \cite{Perez:2012mj}.}
   \label{tab:10}
\end{table} 

With these $\mathcal{S}_3$ charge assignments, the renormalizable Higgs-messenger interactions in the superpotential then take the form
\begin{eqnarray}
W_H= m\mathcal{H}_u \mathcal{H}_d+ \lambda ( X_H \mathcal H_u\mathcal H_d),
\label{higgssuperpotprz}
\end{eqnarray}
in which $\lambda$ is a dimensionless coupling, $m$ is a supersymmetric mass parameter ({\it i.e.}, a $\mu$ term), and the parentheses denote $\mathcal{S}_3$ contractions.  Through some hidden sector dynamics $X_H$ acquires a vacuum expectation value for its scalar and $F$-components, which is parametrized as follows:
\begin{eqnarray}
\langle \lambda X_H \rangle = M  \left (\begin{array}{c} \sin\phi \\ \cos\phi \end{array} \right ) +\theta^2 F  \left (\begin{array}{c} \sin\xi \\ \cos\xi \end{array} \right ),
\label{deltavev}
\end{eqnarray}
where $\phi$ and $\xi$ characterize the vev directions of the scalar and $F$ components, respectively.  Here we will work in the limit in which $F\ll M^2$ for simplicity.  After symmetry breaking, the effective superpotential takes the following form (in self-evident notation):
\begin{eqnarray}
W_H&=& \mathcal{H}_u^T \left (\begin{array}{cc} M \sin\phi & m\\ m & M \cos\phi \end{array} \right ) \mathcal{H}_d+\theta^2\; \mathcal{H}_u^T  \left ( \begin{array}{cc} F\sin\xi & 0\\ 0 & F\cos\xi \end{array} \right ) \mathcal{H}_d \nonumber \\
&\equiv& \mathcal{H}_u^T\left ( \mathbb{M}+\theta^2\; \mathbb{F} \right ) \mathcal{H}_d.
\label{higgseffsuperpotprz}
\end{eqnarray}
As discussed in PRZ \cite{Perez:2012mj}, it is preferable to consider the case that
\begin{eqnarray}
[\mathbb{M}, \mathbb{F}]=0.
\label{commutator}
\end{eqnarray}
Once Eq.~(\ref{commutator}) is imposed, the same unitary rotation diagonalizes both $\mathbb{M}$ and $\mathbb{F}$.  Thus, in the situation of interest, in which there is a mass hierarchy obtained upon the diagonalization of these quantities such that the lighter states can be identified as $H_{u,d}$ and the heavier state as $M_{u,d}$ (note here $N=1$ with this minimal particle content), the heavier states can be smoothly integrated out to obtain the effective theory.   PRZ showed that if $[\mathbb{M}, \mathbb{F}]\neq 0$, the resulting soft mass parameters have some pathologies, including one-loop contributions to the soft scalar mass-squared parameters that are not strongly suppressed in the $F\ll M^2$ limit \cite{Perez:2012mj}.  Thus, we will focus on the case of the vanishing commutator of Eq.~(\ref{commutator}), which yields the condition:
\begin{eqnarray}
[\mathbb{M}, \mathbb{F}]= \left (\begin{array}{cc} 0 & m F(\cos\xi-\sin\xi)\\ -m F(\cos\xi-\sin\xi) &0 \end{array} \right ) = 0.
\label{vanishingcommutator}
\end{eqnarray}
Neglecting the trivial solutions to Eq.~(\ref{vanishingcommutator}) in which $m$ and/or $F$ are equal to zero, we see that we need $\xi=\pi/4$, {\it i.e.}, $\mathbb{F}$ must be proportional to the identity, while $\mathbb{M}$ is not constrained.    

In this case, we see immediately that this scenario suffers from a severe $\mu/B_\mu$ problem.  More precisely, as $\mathbb{F}$ is proportional to the identity, an eigenvalue hierarchy is not possible, and hence $b= B_\mu \mu\sim O(F)$.  While it is in principle possible to obtain a hierarchy of eigenvalues for $\mathbb{M}$ (for example, in PRZ there is an effective seesaw structure for the $\mu$ term that results from taking $\phi=0$ \cite{Perez:2012mj}), generally we have $B_\mu \gg \mu$, and thus if $\mu\sim m_{\rm soft}$, $B_\mu \gg m_{\rm soft}$.  

The $\mu/B_\mu$ problem encountered here is not surprising, given the well-known fact that gauge-mediated models generically suffer from a $\mu/B_\mu$ problem (see e.g.~\cite{Giudice:1998bp} for an overview).  However, the problem here is particularly severe.  To see this more clearly, recall that it has long been realized that a direct superpotential coupling of the supersymmetry breaking field  to the electroweak Higgs doublets $H_{u,d}$ is problematic because it results in an undesirable hierarchy between $\mu$ and $B_\mu$. With the usual notation that the supersymmetry-breaking field is denoted by $X$, the superpotential 
\begin{eqnarray}
W_{\rm H}=\lambda X H_u H_d
\label{muprob}
\end{eqnarray}
generates a tree-level value for both $\mu$ and $b= B_\mu \mu$:
\begin{eqnarray}
\mu= \lambda \langle X \rangle, \qquad b=B_\mu \mu=\lambda \langle F_X \rangle.
\end{eqnarray}
Given that $m_{\rm soft} \sim (1/(16\pi^2)) F_X/X$, we have  
\begin{eqnarray}
B_\mu=\langle F_X\rangle/\langle X \rangle \sim 16\pi^2 m_{\rm soft},
\end{eqnarray}
and hence if $\mu \sim m_{\rm soft}$, $B_\mu$ is too large by a loop factor, which indicates that it is desirable to eliminate the coupling of Eq.~(\ref{muprob}).  In our framework, however, $H_{u,d}$ are embedded together with the doublet messengers $M_{ui,di}$ into $\mathcal{S}_3$ representations. A nonvanishing superpotential coupling between $X$ and the $M_{ui,di}$ then implies that the superpotential coupling of $X$ to the Higgs fields as in Eq.~(\ref{muprob}) is automatically also present, resulting in a severe $\mu/B_\mu$ problem.  

To move forward, therefore, it is necessary to avoid this problematic result.   That being said, most known approaches to resolving the $\mu/B_\mu$ problem of gauge mediation without fine-tuning begin by forbidding the coupling of Eq.~(\ref{muprob}) and generating $\mu$ and $B_\mu$ from alternative operators (see e.g.~\cite{Dvali:1996cu,Giudice:2007ca}).  It is not at all obvious how such approaches could work in our framework.   A  (less ambitious) option is to construct scenarios that alleviate this problem through fine-tuning.  To be more precise, this would entail having a situation in which it is possible to fine-tune both $\mu$ and $B_\mu$ {\it separately}.  This would not be a true solution to the $\mu/B_\mu$ problem of gauge mediation in that tuning is required, but it does allow for the construction of phenomenologically viable models of the soft terms.   This is the approach we will take in this paper.

We see that even with allowing fine-tuning, our minimal $\mathcal{S}_3$ Higgs-messenger scenario given above is not viable, as merely setting the $F$ term component of $X_H$ that couples to the eventual electroweak Higgs doublets to zero is not consistent with the requirement that $[\mathbb{M},\mathbb{F}]=0$.   Hence, to construct a working (toy) model, we need to extend this model to include additional degrees of freedom.  There are several model-building directions that can be taken:
\begin{itemize}
\item One option is to add additional singlet superfields that do not develop $F$ terms and attempt to address the $\mu/B_\mu$ problem via a next-to-minimal supersymmetric standard model (NMSSM)-like approach in which one or more singlets are tied with electroweak symmetry breaking (see e.g.~\cite{Agashe:1997kn,Giudice:1998bp}).

\item A second option is to include an additional supersymmetry breaking field that couples to $\mathcal{H}_{u,d}$.  This additional field, which we will call $X^\prime$, would need to have different $\mathcal{S}_3$ quantum numbers from $X_H$.  The different $\mathcal{S}_3$ contractions of $X^\prime \mathcal{H}_u\mathcal{H}_d$ and $X_H \mathcal{H}_u \mathcal{H}_d$ then will result in a different structure for $\mathbb{M}$ and $\mathbb{F}$, allowing for new possibilities for generating realistic mass hierarchies while satisfying $[\mathbb{M},\mathbb{F}]=0$.   

Indeed, we already have a candidate for this field.  It is $X_T$, the $\mathcal{S}_3$ singlet field that couples to the messenger triplets.  In our minimal scenario, $X_H$ couples only to $\mathcal{H}_{u,d}$ and $X_T$ only couples to the triplets; however, the $\mathcal{S}_3$ assignments certainly allow for $X_T$ to couple to $\mathcal{H}_{u,d}$.

\item A third option is to keep the feature that it is only $X_H$ that couples to the Higgs-messenger fields and  enlarge the Higgs-messenger sector particle content to include different $\mathcal{S}_3$ representations.  In this case, the fields $\mathcal{H}_{u,d}$ as given in Eq.~(\ref{higgs_s3}) then include both doublets and singlets of $\mathcal{S}_3$.  Depending on the details of the mass matrices for these fields, this can result in additional messenger fields or additional electroweak Higgs fields in the theory.  In either case, the additional degrees of freedom gained through using more Higgs-messenger fields provide new possibilities for obtaining viable scenarios in which $\mu$ and $B_\mu$ can be separately tuned while maintaining $[\mathbb{M},\mathbb{F}]=0$.

\end{itemize}
Each of these possibilities lead to intriguing model-building directions.  In the context of the MSSM, the second and third choices are of particular interest.  In this paper, we will focus on the third option, as it turns out to be the most straightforward direction for obtaining models of the MSSM soft terms.  We leave the exploration of the second option to future work.

\section{An extended Higgs-messenger sector}

In this section, we will construct a scenario in which we can separately tune $\mu$ and $B_\mu$  while maintaining $[\mathbb{M},\mathbb{F}]=0$, and determine the resulting messenger Yukawa couplings their subsequent contributions to the MSSM soft terms. As described in the previous section, this scenario will include two supersymmetry breaking fields: $X_H$, which is a $\mathbf{2}$ of $\mathcal{S}_3$ and couples only to the Higgs-messenger fields at the renormalizable level, and $X_T$, which is a $\mathcal{S}_3$ singlet that couples only to the triplet messengers at the renormalizable level.  To this, we add the following Higgs-messenger sector field content.  As in the minimal case also studied by PRZ, we include a pair of messenger-Higgs fields that are in the $\mathbf{2}$ representation of $\mathcal{S}_3$.  We will label these fields by $\mathcal{H}^{(2)}_{u,d}$.  We also include a pair of Higgs-messenger fields that are $\mathcal{S}_3$ singlets, which we will denote by $\mathcal{H}^{(1)}_{u,d}$.  These charge assignments are shown in Table~\ref{tab:11} \footnote{We assume throughout that $X_T$ and the triplet messengers, which are $\mathcal{S}_3$ singlets, do not have renormalizable couplings to the Higgs-messenger sector.  In practice, this requires the addition of additional symmetries to forbid such couplings, which does not pose a significant challenge to arrange.}.
\begin{table}[htbp]
   \centering
    \begin{tabular}{c|cccc|c}
     & $\Hu{\mathbf 2}$&$\Hu{\mathbf 1}$ & $\Hd{\mathbf 2}$& $\Hd{\mathbf 1}$  &$X_H$\\
    \hline
    $\mathcal{S}_3$ &$\mathbf 2 $& $\mathbf 1$& $\mathbf 2 $& $\mathbf 1 $  &\textbf 2\\
   \end{tabular}
   \caption{The $\mathcal{S}_3$ charges for the extended Higgs-messenger model described in this section.}
   \label{tab:11}
\end{table} 
The renormalizable superpotential couplings of $X_H$ to the Higgs-messenger fields then take the form
\begin{eqnarray}
W_H&=& \lambda(X_H\mathcal{H}^{(2)}_u \mathcal{H}^{(2)}_d)+\lambda'(X_H \mathcal{H}^{(1)}_u \mathcal{H}^{(2)}_d)+\lambda''(X_H \mathcal{H}^{(2)}_u \mathcal{H}^{(1)}_d) \\
&=& M \mathcal H_u^T\left(\begin{matrix}\sin\phi&0 &\epsilon'\cos\phi\\ 0 &\cos\phi &\epsilon'\sin\phi\\\epsilon''\cos\phi &\epsilon'' \sin\phi &0
\end{matrix}\right)\mathcal H_d  + \theta^2 F \mathcal H_u^T\left(\begin{matrix}\sin\xi&0&\epsilon'\cos\xi\\ 0 &\cos\xi &\epsilon'\sin\xi\\\epsilon''\cos\xi &\epsilon'' \sin\xi & 0 \end{matrix}\right)\mathcal H_d, \nonumber
\label{extendedS3}
\end{eqnarray}
in which $\epsilon'=\lambda'/\lambda$, $\epsilon''=\lambda''/\lambda$, and the quantities $\mathcal{H}_{u,d}$ are now given by
\begin{eqnarray}
\mathcal{H}_u = \left (\begin{array}{c} (\mathcal{H}^{(2)}_u)_1 \\ (\mathcal{H}^{(2)}_u)_2 \\ \mathcal{H}^{(1)}_u \end{array} \right ), \qquad \mathcal{H}_d = \left (\begin{array}{c} (\mathcal{H}^{(2)}_d)_1 \\ (\mathcal{H}^{(2)}_d)_2 \\ \mathcal{H}^{(1)}_d \end{array} \right ).
\end{eqnarray}
Let us assume for the moment that there are no bare mass terms.  With only the couplings of Eq.(\ref{extendedS3}), the 
the commutation condition $[\mathbb M,\mathbb F]=0$ only has solutions when $\epsilon'=\epsilon''=0$ or $\phi=\xi$.  The case with $\epsilon'=\epsilon''=0$ results in uncoupled singlets that do not mix with the $\mathcal S_3$ doublets; and in the case that $\phi=\xi$, the two matrices have identical structure and thus the eigenvalues will be proportional, resulting again in the $\mu/B_\mu$ problem that $B_\mu/\mu =F/M$.

Hence, bare mass terms are needed, and indeed they are allowed by the $\mathcal{S}_3$ symmetry.  Including them results in the following modification to Eq.~(\ref{extendedS3}):
\begin{eqnarray}
W_H= \lambda(X_H\mathcal{H}^{(2)}_u \mathcal{H}^{(2)}_d)+\lambda'(X_H \mathcal{H}^{(1)}_u \mathcal{H}^{(2)}_d)+\lambda''(X_H \mathcal{H}^{(2)}_u \mathcal{H}^{(1)}_d) +  \kappa M (\mathcal{H}^{(2)}_u \mathcal{H}^{(2)}_d)+  \kappa^\prime M (\mathcal{H}^{(1)}_u \mathcal{H}^{(1)}_d),\nonumber \\
\end{eqnarray}
such that in this case, $\mathbb{M}$ and $\mathbb{F}$ are given by
\begin{eqnarray}
\mathbb{M}= M \left(\begin{matrix}\sin\phi&\kappa &\epsilon'\cos\phi\\ \kappa &\cos\phi &\epsilon'\sin\phi\\\epsilon''\cos\phi &\epsilon'' \sin\phi &\kappa^\prime \end{matrix}\right), 
\qquad  \mathbb{F}=F \left(\begin{matrix}\sin\xi&0&\epsilon'\cos\xi\\ 0 &\cos\xi &\epsilon'\sin\xi\\\epsilon''\cos\xi &\epsilon'' \sin\xi & 0 \end{matrix}\right). \nonumber
\end{eqnarray}
For simplicity, we will take the case that $\epsilon'' =\epsilon$, such that $\mathbb{M}$ and $\mathbb{F}$ are symmetric matrices, and further take $\epsilon'=1$ for concreteness.  We will also restrict ourselves to real quantities only.  In this case, the commutation condition results in the nontrivial solution
\begin{eqnarray}
\kappa^\prime = \kappa = \frac{\sin(\phi-\xi)}{\cos \xi-\sin\xi},
\end{eqnarray}
which clearly only holds for $\xi\neq \pi/4$, whereas for $\xi=\pi/4$, the only solution is $\phi=\pi/4$, with no constraints on $\kappa$.  For reasons that will become clear shortly, we will focus on the solution that is valid for $\xi\neq \pi/4$.  In this case, we have
\begin{eqnarray}
\mathbb{M} =M \cos \phi \left (\begin{array}{ccc} \tan\phi & \frac{\tan \phi-\tan \xi}{1-\tan \xi} & 1 \\   \frac{\tan \phi-\tan \xi}{1-\tan \xi}& 1 & \tan \phi \\ 1 & \tan \phi &  \frac{\tan \phi-\tan \xi}{1-\tan \xi} \end{array} \right ), \qquad \mathbb{F} = F \cos \xi \left (\begin{array}{ccc} \tan \xi &0&1 \\ 0&1 & \tan \xi \\ 1& \tan \xi & 0 \end{array} \right ).
\end{eqnarray}
Since $\mathbb{M}$ and $\mathbb{F}$ are simultaneously diagonalizable, we see from the form of $\mathbb{F}$ that the mixing only depends on $\tan \xi$.  The eigenvalues of $\mathbb{F}$, which we denote by $F_{i=1,2,3}$ take the form
\begin{eqnarray}
F_1= F (\cos\xi+\sin \xi), \qquad F_{2,3} = \mp F\sqrt{1-\sin\xi \cos\xi}.
\label{fevals}
\end{eqnarray}
Similarly, the eigenvalues of $\mathbb{M}$, denoted by $M_{i=1,2,3}$, are given by
\begin{eqnarray}
M_1=M \left ( \frac{\cos(\xi+\phi)-2 \sin(\xi-\phi)}{\cos\xi-\sin\xi} \right ), \qquad M_{2,3} = \mp M \left ( \frac{\cos\phi-\sin\phi}{\cos\xi-\sin\xi} \right ) \sqrt{1-\sin\xi \cos\xi}.
\label{mevals}
\end{eqnarray}
The eigenvalues $F_1$ and $M_1$ are associated with the trimaximal vector, $(1/\sqrt{3} )(1,1,1)$.
The minus sign associated with $F_2$ and $M_2$  can be removed by a rephasing of its associated eigenvector in either $\mathcal{R}_u$ or $\mathcal{R}_d$.  With this in mind, one specific choice for the rotation matrices $\mathcal{R}_{u,d}$ is as follows:
\begin{eqnarray}
\mathcal{R}_{u}= \left ( \begin{array}{ccc} \frac{1}{\sqrt{3}} & \mp \frac{1}{\sqrt{2}N_2} \left (1-\frac{\tan\xi}{1+\delta} \right ) & -\frac{1}{\sqrt{2}N_3} \left (1-\frac{\tan\xi}{1-\delta} \right )\\  
\frac{1}{\sqrt{3}} & \mp \frac{1}{\sqrt{2}N_2} \frac{\tan\xi}{1+\delta} & -\frac{1}{\sqrt{2}N_3} \frac{\tan\xi}{1-\delta}\\ 
\frac{1}{\sqrt{3}} & \pm \frac{1}{\sqrt{2}N_2} & -\frac{1}{\sqrt{2}N_3} \end{array} \right )
\end{eqnarray}
in which the upper (lower) sign in the second column denotes $\mathcal{R}_u$ ($\mathcal{R}_d$), 
$\delta = \sqrt{1-\tan\xi+\tan\xi^2} = \sqrt{1-\sin\xi\cos\xi}/\cos\xi$, and the coefficients $N_{2,3}$ take the form
\begin{eqnarray}
N_2 &=& \sqrt{1-\frac{\tan\xi}{1+\delta}+\left (\frac{\tan\xi}{1+\delta} \right )^2}\\
N_3 &=&\sqrt{1-\frac{\tan\xi}{1-\delta}+\left (\frac{\tan\xi^2}{1-\delta}\right )^2}.
\end{eqnarray}
We now need to build in the eigenvalue hierarchies, {\it i.e.}, $F_1 \equiv b \ll F_{2,3}$ and $M_1\equiv \mu \ll M_{2,3}$. We start by setting $\mu=M_1$ and $b = F_1$, such that we obtain one light pair of doublets that will be identified as $H_{u,d}$.   Eq.~(\ref{fevals}) shows that $b$ is naturally $O(F)$, but a light eigenvalue can be obtained for $\xi \rightarrow -\pi/4$, with $b= 0$ in the exact limit that $\xi= -\pi/4$.  Writing  $\xi=-\pi/4+\eta$, we have
\begin{eqnarray}
\frac{b}{F} \equiv F_1 = \sqrt{2} \eta +O(\eta^2), \qquad \frac{F_{2,3}}{F} = \sqrt{\frac{3}{2}}+O(\eta^2).
\end{eqnarray}
Hence, a tuning of $b$ to phenomenologically acceptable values can be done via the parameter $\eta$.  Turning to the issue of tuning the $\mu$ parameter, we see that if $\phi=\xi$, then $\mu/M = b/F$, which is the statement of the $\mu/B_\mu$ problem in gauge mediation.  Therefore, a detuning of $\phi$ from $\xi $ is needed, while still obtaining $\mu \ll M_{2,3} \sim O(M)$.  Setting $\phi = \xi+\rho$, we have to leading order that in the $\xi\rightarrow -\pi/4$ limit,
\begin{eqnarray}
\frac{\mu}{M}\simeq \sqrt{2} \eta +\frac{3}{\sqrt{2}}\rho, \qquad \frac{M_{2,3}}{M} \simeq \sqrt{\frac{3}{2}},
\end{eqnarray}
which demonstrates that in the expression for $\mu$, the term proportional to $\rho$ must be able to counter the $\sqrt{2}\eta$ term sufficiently, such that the appropriate hierarchy between $\mu$ and $B_\mu$ can be achieved.  More precisely, what is needed is that in this limit,
\begin{eqnarray}
B_\mu = \frac{b}{\mu} =\frac{F}{M} \frac{2\eta}{3\rho} \sim \frac{1}{16\pi^2} \frac{F}{M} \sim m_{\rm soft},
\end{eqnarray} 
{\it i.e.}, both $\eta\ll 1$ and $\rho \ll 1$, and $\rho/\eta \rightarrow \sim (4 \pi)^2$.  In the $\xi\rightarrow -\pi/4$ limit, the matrices $\mathcal{R}_{u,d}$ are
\begin{eqnarray}
\mathcal{R}_{u,d}= \left ( \begin{array}{ccc} \frac{1}{\sqrt{3}} & \mp \frac{1}{2} \left (1+\frac{1}{\sqrt{3}} \right) & \frac{1}{2} \left (1-\frac{1}{\sqrt{3}} \right) \\  \frac{1}{\sqrt{3}} & \pm \frac{1}{2} \left (1-\frac{1}{\sqrt{3}} \right) & -\frac{1}{2} \left (1+\frac{1}{\sqrt{3}} \right) 
\\  \frac{1}{\sqrt{3}} &  \pm \frac{1}{\sqrt{3}} &  \frac{1}{\sqrt{3}} \end{array} \right )+ O(\eta).
\label{rotationmatrices}
\end{eqnarray}
Though this fine-tuning is not esthetically very pleasing, it is worth noting that something has been accomplished in this section: we are now able to construct viable models since $\mu$ and $b$ can be tuned {\it independently} while keeping $[\mathbb{M}, \mathbb{F}]=0$, which was not possible in the more minimal scenario described previously.  Thus, the $\mu/B_\mu$ problem has been alleviated, though not solved dynamically.

Hence, this Higgs-messenger sector will be our starting point for model-building.  To this sector, we will add two pairs of $SU(3)$ messenger triplets to preserve anomaly cancellation, gauge coupling, unification, and generate a nonzero gluino mass.   With this result in hand, we now turn to an examination of the possibilities for generating the Yukawa couplings to the observable sector and the resulting gauge-mediated MSSM soft terms.

\section{Models}

We now turn to the observable sector, and discuss options for embedding the MSSM matter fields into representations of $\mathcal{G}$.  There are clearly a variety of possibilities.  If the MSSM matter fields have nontrivial $\mathcal{G}$ quantum numbers, then by definition $\mathcal{G}$ then is (at least part) of the family symmetry group as well as the Higgs-messenger symmetry group.  This is the strategy that was employed by PRZ \cite{Perez:2012mj}; working with $\mathcal{G}=\mathcal{S}_3$ (as we do here), they employed the minimal Higgs-messenger sector of Section~\ref{backgroundsection} and considered a two-family scenario in which all MSSM matter fields were assigned to $\mathbf{2}$'s of $\mathcal{S}_3$.  Hence, one possibility for us to explore is to extend this to three families, using the modified Higgs-messenger sector of the previous section.  At the other end of the spectrum, another possibility is to make the MSSM matter fields inert with respect to $\mathcal{S}_3$.   There are also mixed scenarios in which some of the MSSM states are $\mathcal{S}_3$ singlets and others are not.  We will not attempt to be systematic and classify all scenarios in this work, but rather focus on a few representative yet simple examples.

An important model-building requirement is that it is desirable to have the top quark Yukawa coupling to arise from a renormalizable operator.   In general, this means that if the Higgs-messenger sector consists only of nontrivial representations of $\mathcal{G}$ ({\it i.e.}, it has no $\mathcal{G}$ singlets), at least some of what would be identified as the top quark degrees of freedom would also need to be in nontrivial representations of $\mathcal{G}$, otherwise $\mathcal{G}$ would forbid a renormalizable top quark coupling.   However, if the Higgs-messenger sector includes $\mathcal{G}$ singlets, the top quark degrees of freedom can remain inert.

In the $\mathcal{S}_3$ models considered here, the analogous situation is that if the Higgs-messenger field content only includes doublets, we would need to have either $Q_3$ or $\overline{u}_3$ (or both) as components of $\mathcal{S}_3$ doublets.   However, as demonstrated previously, one of the ways to alleviate the $\mu/B_\mu$ problem is to include Higgs-messenger fields that are $\mathcal{S}_3$ singlets in addition to the Higgs-messenger $\mathcal{S}_3$ doublets.  Therefore, our Higgs-messenger sector allows for several situations in which we obtain a renormalizable top quark Yukawa coupling.  We can either have this coupling originate from the coupling to the $\mathcal{S}_3$ singlet, $\mathcal{H}^{(1)}_u$, in which case the top quark degrees of freedom are $\mathcal{S}_3$ singlets, or we can have the top quark Yukawa coupling arise from the coupling to the $\mathcal{S}_3$ doublet, $\mathcal{H}^{(2)}_u$, in which case some or all of the top quark degrees of freedom (together with other quark degrees of freedom) should be embedded in $\mathcal{S}_3$ doublets.  We will find it useful in what follows to consider these two different categories of models separately.

\subsection{Top quark Yukawa coupling from $\mathcal{H}^{(1)}_u$ models}

In this set of models, the top quark degrees of freedom are inert with respect to $\mathcal{S}_3$, such that 
\begin{eqnarray}
W_{\rm top} = y_t Q_3 \overline{u}_3 \mathcal{H}^{(1)}_u,
\end{eqnarray}
in which $y_t$ is an $O(1)$ number.  Using our result for $\mathcal{R}_u$ in the $\xi\rightarrow -\pi/4$ limit as given in Eq.~(\ref{rotationmatrices}), 
\begin{eqnarray}
\mathcal{H}^{(1)}_u =  \left [ \mathcal{R}_u \left ( \begin{array}{c} H_u \\ M_{u1} \\ M_{u2} \end{array} \right ) \right ]_3 = \frac{1}{\sqrt{3}}(H_u+M_{u1}+M_{u2} ),
\label{hu1}
\end{eqnarray}
we obtain equal values for the leading order contributions to the MSSM top quark Yukawa coupling and the messenger top quark Yukawa couplings,
\begin{eqnarray}
W_{\rm top} = Y_t Q_3 \overline{u}_3 H_u+ Y_t Q_3 \overline{u}_3 M_{u1}+Y_t Q_3 \overline{u}_3 M_{u2},
\label{model1top}
\end{eqnarray}
in which $Y_t=y_t/\sqrt{3}$.  Thus, the messenger Yukawa couplings $Y^\prime_{t1}$ and $Y^\prime_{t2}$ of the top quark to the messengers $M_{u1}$ and $M_{u2}$ are both equal to the top quark Yukawa coupling $Y_t$.

Focusing for simplicity on third family Yukawa couplings only, we can have the $b$ and $\tau$ Yukawas either from similar operators, or they can in principle arise from nonrenormalizable operators (or at least, these are the dominant contributions).  If the $b$  and $\tau$ degrees of freedom are also $\mathcal{S}_3$ singlets, the Yukawa couplings arise from the following superpotential, which we will label as $W_{\rm A 1}$:
\begin{eqnarray}
W_{\rm A 1} = y_t Q_3 \overline{u}_3 \mathcal{H}^{(1)}_u+ y_b Q_3 \overline{d}_3 \mathcal{H}^{(1)}_d+y_\tau L_3 \overline{e}_3 \mathcal{H}^{(1)}_d,
\end{eqnarray}
in self-evident notation.Since we have, in analogy with Eq.~(\ref{hu1}), 
\begin{eqnarray}
\mathcal{H}^{(1)}_d =  \left [ \mathcal{R}_d \left ( \begin{array}{c} H_d \\ M_{d1} \\ M_{d2} \end{array} \right ) \right ]_3 = \frac{1}{\sqrt{3}}(H_d-M_{d1}+M_{d2} ),
\label{hd1}
\end{eqnarray}
the Yukawa interactions of the third generation fields with the Higgs and messengers take the form
\begin{eqnarray}
W_{\rm A 1}&=&Y_t Q_3 \overline{u}_3 H_u+ Y_t Q_3 \overline{u}_3 M_{u1}+Y_t Q_3 \overline{u}_3 M_{u2} + Y_b Q_3 \overline{d}_3 H_d- Y_b Q_3 \overline{d}_3 M_{d1}+Y_b Q_3 \overline{d}_3 M_{d2} \nonumber \\ &+& Y_\tau L_3 \overline{e}_3 H_d- Y_\tau L_3 \overline{e}_3 M_{d1}+Y_\tau L_3 \overline{e}_3 M_{d2},
\label{a1superpot}
\end{eqnarray}
in which $Y_b=y_b/\sqrt{3}$ and $Y_\tau=y_\tau/\sqrt{3}$, such that the magnitudes of the $b$ and $\tau$ messenger Yukawa couplings are thus also  identical to their MSSM counterparts.  We will refer to this scenario as Model A1.    In this scenario, the corrections to the MSSM soft terms due to the messenger Yukawas are as follows:

\begin{eqnarray}
\delta m^2_{Q_{33}}&=& \frac{\Lambda^2}{(4\pi)^4} \left [  36 (Y_t^4+Y_b^4)+ 8 Y_b^2 (Y_t^2+ Y_\tau^2) -2 \tilde{g}_u^2 Y_t^2-2 \tilde{g}_d^2 Y_b^2 \right ] \nonumber \\
\delta m^2_{\bar{u}_{33}}&=& \frac{\Lambda^2}{(4\pi)^4} \left [  72 Y_t^4+8 Y_t^2 Y_b^2 -4 \tilde{g}_u^2 Y_t^2 \right ], \;\; 
\delta m^2_{\bar{d}_{33}}=
\frac{\Lambda^2}{(4\pi)^4} \left [  72 Y_b^4+8 Y_b^2 Y_t^2 +16 Y_b^2 Y_\tau^2-4 \tilde{g}_d^2 Y_b^2\right ]\nonumber \\
\delta m^2_{L_{33}}&=& \frac{\Lambda^2}{(4\pi)^4} \left [  20Y_\tau^4+ 24 Y_b^2 Y_\tau^2 -2 \tilde{g}_e^2 Y_\tau^2 \right ], \;\;\;  
\delta m^2_{\bar{e}_{33}}= 
\frac{\Lambda^2}{(4\pi)^4} \left [  40 Y_\tau^4+48 Y_b^2 Y_\tau^2-4 \tilde{g}_e^2 Y_\tau^2 \right ]\nonumber \\
\delta m^2_{H_u} &=& \frac{\Lambda^2}{(4\pi)^4} \left [ -18 Y_t^4 - 6 Y_b^2 Y_t^2 \right ], \;\;\; 
\delta m^2_{H_d} = 
\frac{\Lambda^2}{(4\pi)^4} \left [ -18 Y_b^4 - 6 Y_b^2 Y_t^2-6 Y_\tau^4 \right ]\nonumber \\
\tilde{A}_{u_{33}}&=& -\frac{\Lambda}{(4\pi)^2} \left [6 Y_t^3+2 Y_b^2 Y_t \right ], \;\;
\tilde{A}_{d_{33}}=
-\frac{\Lambda}{(4\pi)^2} \left [6 Y_b^3+2 Y_t^2 Y_b \right ], \;\; 
\tilde{A}_{e_{33}}=
-\frac{\Lambda}{(4\pi)^2} \left [ 6 Y_\tau^3 \right ],
\label{softtermsA1}
\end{eqnarray}
in which
\begin{eqnarray} 
\tilde{g}_u^2 = \frac{16}{3} g_3^2+3 g_2^2 +\frac{13}{15} g_1^2, \qquad \tilde{g}_d^2 = \frac{16}{3} g_3^2+3 g_2^2 + \frac{7}{15} g_1^2,\qquad \tilde{g}_e^2 = 3 g_2^2 +\frac{9}{5} g_1^2,
\end{eqnarray}
$Y_{t,b,\tau}$ denote the MSSM Yukawa couplings, and $\Lambda = \vert F_{2,3}/M_{2,3} \vert$ \footnote{Here the $\tilde{A}$ terms correspond to the trilinear scalar couplings in the Lagrangian, {\it i.e.} $-\mathcal{L} \sim \tilde{A}_{ijk} \phi_i \phi_j \phi_k$.}.  The full expressions for the soft terms also include the standard gauge-mediated contributions (see e.g.~\cite{gauge1,gauge2,gauge3,Giudice:1998bp}).

Alternatively, we can envision scenarios in which the $b$ and $\tau$ degrees of freedom are $\mathcal{S}_3$ singlets, but couple only to $\mathcal{H}^{(2)}_d$ via nonrenormalizable operators.  Such operators require the introduction of additional degrees of freedom that we will denote collectively by $\varphi$ (in practice, there could be a set of fields $\varphi_i$), which are $\mathbf{2}$'s of $\mathcal{S}_3$.   We will refer to this scenario as Model A2.
In this case, the following superpotential Yukawa interactions are allowed by the gauge and $\mathcal{S}_3$ symmetries:
\begin{eqnarray}
W_{\rm A 2} = y_t Q_3 \overline{u}_3 \mathcal{H}^{(1)}_u+ \frac{1}{\tilde{\Lambda}} \tilde{Y}_b Q_3 \overline{b}_3 (\varphi \mathcal{H}^{(2)}_d ) + \frac{1}{\tilde{\Lambda}} \tilde{Y}_\tau L_3 \overline{e}_3 (\varphi \mathcal{H}^{(2)}_d )+\ldots,
\end{eqnarray}
in which $\tilde{\Lambda}$ is a (presumably high) scale, $\tilde{Y}_{b,\tau}$ are $O(1)$ factors, and we have neglected subdominant interactions. The detailed couplings depend on the vacuum expectation value of $\varphi$.  More explicitly, 
\begin{eqnarray}
W_{\rm A 2} = \tilde{y}_t Q_3 \overline{u}_3 \mathcal{H}^{(1)}_u+ \frac{1}{\tilde{\Lambda}} \tilde{Y}_b Q_3 \overline{b}_3 (\varphi_1 \mathcal{H}^{(2)}_{d2}+\varphi_2 \mathcal{H}^{(2)}_{d1}) + \frac{1}{\tilde{\Lambda}} \tilde{Y}_\tau L_3 \overline{e}_3  (\varphi_1 \mathcal{H}^{(2)}_{d2}+\varphi_2 \mathcal{H}^{(2)}_{d1}) +\ldots.
\end{eqnarray}
Given that in our scenario for $\xi = -\pi/4$, the components of $\mathcal{H}^{(2)}_d$ are given by
\begin{eqnarray}
\mathcal{H}^{(2)}_d=\left (\begin{array}{c} \mathcal{H}^{(2)}_{d1}\\ \mathcal{H}^{(2)}_{d2} \end{array} \right ) =\left (\begin{array}{c}  \frac{1}{\sqrt{3}} H_d + \frac{1}{2} \left (1+\frac{1}{\sqrt{3}} \right ) M_{d1}+\frac{1}{2} \left (1-\frac{1}{\sqrt{3}} \right ) M_{d2} \\ \frac{1}{\sqrt{3}} H_d + \frac{1}{2} \left (-1+\frac{1}{\sqrt{3}} \right ) M_{d1}-\frac{1}{2} \left (1+\frac{1}{\sqrt{3}} \right ) M_{d2} \end{array} \right ),
\end{eqnarray}
the expression for $W_{\rm A 2}$ takes the form
\begin{eqnarray}
W_{\rm A 2}&=& Y_t Q_3 \overline{u}_3 H_u+ Y_t Q_3 \overline{u}_3 M_{u1}+Y_t Q_3 \overline{u}_3 M_{u2}
+Y_bQ_3 \overline{b}_3 H_d+ Y^\prime_{b1}Q_3 \overline{b}_3 M_{d1}+Y^\prime_{b2}Q_3 \overline{b}_3 M_{d2}
 \nonumber \\ &+& Y_\tau L_3 \overline{e}_3 H_d+ Y^\prime_{\tau 1} L_3 \overline{e}_3 M_{d1}+Y^\prime_{\tau 2} L_3 \overline{e}_3 M_{d2},
\end{eqnarray}
in which the MSSM and messenger Yukawas are given by
\begin{eqnarray}
Y^\prime_{t1} &=& Y^\prime_{t2}= Y_t, \qquad Y_b = \frac{\tilde{Y}_b}{\sqrt{3}\tilde{\Lambda}} (\varphi_1+\varphi_2), \qquad Y_\tau = \frac{\tilde{Y}_\tau}{\sqrt{3}\tilde{\Lambda}} (\varphi_1+\varphi_2), \nonumber \\
Y^\prime_{b,\tau1} &=&  \frac{\tilde{Y}_{b,\tau}}{2\tilde{\Lambda}} ((-1+1/\sqrt{3}) \varphi_1+ (1+1/\sqrt{3})\varphi_2),\nonumber \\ 
Y^\prime_{b,\tau 2} &=&  -\frac{\tilde{Y}_{b,\tau}}{2\tilde{\Lambda}} ((1+1/\sqrt{3}) \varphi_1-(1-1/\sqrt{3})\varphi_2 ).
\label{a2superpot}
\end{eqnarray}
Clearly, the $b$ and $\tau$ messenger Yukawas depend on the $\varphi$ direction, and generically are similar in size to their SM counterparts.  However, there are points of interest in which there are exact or near-cancellations, such that these conclusions no longer hold.  For example, we note that if $\varphi_1=-\varphi_2$, the MSSM Yukawas $Y_b$ and $Y_\tau$ are zero, and the messenger Yukawas are equal ($Y_{b1}^\prime = Y_{b2}^\prime$, $Y_{\tau 1}^\prime = Y_{\tau 2}^\prime$).  Another situation of interest occurs in the case that $\varphi_1=(2\pm \sqrt{3})\phi_2$.  In this limit, either $Y_{b1}^\prime$ and $Y_{\tau1}^\prime$ vanish (plus sign) or  $Y_{b2}^\prime$ and $Y_{\tau2}^\prime$ vanish (minus sign), such that the $b$ and $\tau$ couplings are given by
\begin{eqnarray}
Y_{b,\tau}&=&\tilde{Y}_{b,\tau} \frac{1+\sqrt{3}}{2} \frac{\varphi_2}{\tilde{\Lambda}}=-Y^\prime_{b,\tau 2}, \;\; Y^\prime_{b,\tau 1}=0, \qquad (\varphi_1\rightarrow (2+\sqrt{3})\varphi_2) \nonumber \\
Y_{b,\tau}&=&\tilde{Y}_{b,\tau} \frac{-1+\sqrt{3}}{2} \frac{\varphi_2}{\tilde{\Lambda}}=Y^\prime_{b,\tau 1}, \;\; Y^\prime_{b,\tau 2}=0, \qquad (\varphi_1\rightarrow (2-\sqrt{3})\phi_2).
\label{a2specialcases}
\end{eqnarray}
For concreteness, we will focus here on the two simpler cases of Eq.~(\ref{a2specialcases}), which will yield identical phenomenology.  In this situation, the corrections to the soft terms are given by
\begin{eqnarray}
\delta m^2_{Q_{33}}&=& \frac{\Lambda^2}{(4\pi)^4} \left [  36 Y_t^4+12 Y_b^4+ 4 Y_b^2 Y_t^2+ 3 Y_b^2 Y_\tau^2 - 2\tilde{g}_u^2 Y_t^2- \tilde{g}_d^2 Y_b^2 \right ] \nonumber \\
\delta m^2_{\bar{u}_{33}}&=& \frac{\Lambda^2}{(4\pi)^4} \left [  72 Y_t^4+6 Y_t^2 Y_b^2-4 \tilde{g}_u^2  Y_t^2 \right ], \;\;\;\;\;
\delta m^2_{\bar{d}_{33}}= 
\frac{\Lambda^2}{(4\pi)^4} \left [  24 Y_b^4+2 Y_b^2 Y_t^2 + 6 Y_b^2 Y_\tau^2-2  \tilde{g}_d^2 Y_b^2 \right ]\nonumber \\
\delta m^2_{L_{33}}&=& \frac{\Lambda^2}{(4\pi)^4} \left [  6 Y_\tau^4+ 9 Y_b^2 Y_\tau^2 - \tilde{g}_e^2 Y_\tau^2 \right ], \;\; \;\;\;\;
\delta m^2_{\bar{e}_{33}}=
\frac{\Lambda^2}{(4\pi)^4} \left [  12 Y_\tau^4+18 Y_b^2 Y_\tau^2-2 \tilde{g}_e^2 Y_\tau^2 \right ]\nonumber \\
\delta m^2_{H_u} &=& \frac{\Lambda^2}{(4\pi)^4} \left [ -18 Y_t^4 - 3 Y_b^2 Y_t^2 \right ], \;\;\;\;\; 
\delta m^2_{H_d} = 
\frac{\Lambda^2}{(4\pi)^4} \left [ -9 Y_b^4 - 6 Y_b^2 Y_t^2-3 Y_\tau^4 \right ]\nonumber \\
\tilde{A}_{u_{33}}&=& -\frac{\Lambda}{(4\pi)^2} \left [6 Y_t^3+Y_b^2 Y_t \right ], \;\; 
\tilde{A}_{d_{33}}=
-\frac{\Lambda}{(4\pi)^2} \left [3 Y_b^3+2 Y_t^2 Y_b \right ], \;\; 
\tilde{A}_{e_{33}}=
-\frac{\Lambda}{(4\pi)^2}3 Y_\tau^3.
\label{softtermsA2}
\end{eqnarray}
We see that the corrections of Eq.~(\ref{softtermsA2}) are very similar to the case of Model A1 as given in Eq.~(\ref{softtermsA1}), but with smaller corrections in the $b$ and $\tau$ sectors, as expected.

\subsection{Top quark Yukawa coupling from $\mathcal{H}^{(2)}_u$ models}

In this set of models, we must embed at least the top quark degrees of freedom into nontrivial multiplets of $\mathcal{S}_3$ to obtain a renormalizable top quark Yukawa coupling.  The question of the extent to which the remaining MSSM matter fields are also charged under $\mathcal{S}_3$ is a model-building issue;  we will for concreteness focus on a scenario in which the matter fields are embedded in both singlet and doublet representations of $\mathcal{S}_3$, as shown in Table \ref{tab:12}.  Here we note that additional symmetries will in general need to be introduced to prevent additional messenger-matter couplings, but this does not provide a significant model-building challenge. For conciseness, we do not display these constraints explicitly.
\begin{table}[htbp]
   \centering
    \begin{tabular}{c|cccc|cccccccccc|c}
     & $\Hu{\mathbf 2}$&$\Hu{\mathbf 1}$ & $\Hd{\mathbf 2}$& $\Hd{\mathbf 1}$  & $Q_{\mathbf 2}$ &$Q_{\mathbf 1}$& $\bar u_{\mathbf 2}$ & $\bar u_{\mathbf 1 }$&$\bar d_{\mathbf 2}$& $\bar d_{\mathbf 1}$ & $L_{\mathbf{2}}$ & $L_{\mathbf{1}}$ & $\bar{e}_{\mathbf{2}}$ & $\bar{e}_{\mathbf{1}}$&$X_H$\\
    \hline
    $\mathcal{S}_3$ &$\mathbf 2 $& $\mathbf 1$& $\mathbf 2 $& $\mathbf 1 $  & $\mathbf 2 $&$\mathbf 1$  & $\mathbf 2 $&$\mathbf 1$  & $\mathbf 2 $&$\mathbf 1$ & $\mathbf 2 $&$\mathbf 1$  & $\mathbf 2 $&$\mathbf 1$ &\textbf 2\\
   \end{tabular}
   \caption{Charges for an $\st$ model of the Higgs-messenger fields and the MSSM matter fields.}
   \label{tab:12}
\end{table} 
With this set of charge assignments, we see that without further restrictions on the theory, we can have couplings of each to both $\mathcal{H}^{(2)}_{u,d}$ and $\mathcal{H}^{(1)}_{u,d}$.  For example, in the up quark sector, we have 
\begin{eqnarray}
W^{(u)}_{\rm B}= y_u\big[(Q_{\mathbf 2}  \bar u_{\mathbf 2}  \mathcal{H}^{(2)}_u)+\beta_1 (Q_{\mathbf 2}  \bar u_{\mathbf 1}  \mathcal{H}^{(2)}_u)+ \beta_2(Q_{\mathbf 2}  \bar u_{\mathbf 2} \mathcal{H}^{(1)}_u) +\beta_3 (Q_{\mathbf 1}  \bar u_{\mathbf 2}  \mathcal{H}^{(2)}_u)+ \beta_4(Q_{\mathbf 1}  \bar u_{\mathbf 1}  \mathcal{H}^{(1)}_u)\big],
\label{wub}
\end{eqnarray}
in which the $\beta_i$ are arbitrary coefficients in the absence of further model structure (different UV completions may of course result in specific relations between some or all of these coefficients).  Using the notation that $Q=(Q_{\mathbf 2} ,Q_{\mathbf 1})^T$ and $\overline{u}=(\overline{u}_{\mathbf 2} ,\overline{u}_{\mathbf 1})^T$, we can write $W^{(u)}_{\rm B}$ in matrix form as
\begin{eqnarray}
W^{(u)}_{\rm B}=y_uQ^T\left( \begin{matrix} \mathcal{H}^{(2)}_{u1}&\beta_1\mathcal{H}^{(1)}_{u}&\beta_2 \mathcal{H}^{(2)}_{u2}\\ \beta_1 \mathcal{H}^{(1)}_u& \mathcal{H}^{(2)}_{u2}& \beta_2\mathcal{H}^{(2)}_{u1}\\ \beta_3\mathcal{H}^{(2)}_{u2}& \beta_3\mathcal{H}^{(2)}_{u1}&\beta_4 \mathcal{H}^{(1)}_u\end{matrix}\right)\bar u. \label{UpYukawas}
\end{eqnarray}
The analogous quantities $W^{(d)}_{\rm B}$, $W^{(e)}_{\rm B}$ for the down quark and the charged leptons would have similar structure. Depending on model details, their couplings can be suppressed by a Frogatt-Nielsen mechanism. For the sake of simplicity, here we neglect such considerations, as well as the question of the origin of neutrino masses.

To proceed, we need to specify the coefficients in  Eq.~(\ref{UpYukawas}). In this work, we will focus for simplicity on the extremely simple (though contrived) situation  
in which all the coefficients are equal and set to unity.  We will label this scenario as Model B1.  In this case, we have
\begin{eqnarray}
W^{(u)}_{\rm B1}&=&y_uQ^T\left( \begin{matrix} \Hu{\mathbf21}&\Hu{\mathbf1}&\Hu{\mathbf22}\\ \Hu{\mathbf1}& \Hu{\mathbf22}& \Hu{\mathbf21}\\ \Hu{\mathbf22}& \Hu{\mathbf21}&\Hu{\mathbf1}\end{matrix}\right)\bar u.  \label{allsame}
\end{eqnarray}
In terms of the Higgs and messenger mass eigenstates, for $\xi=-\pi/4$ we obtain
\begin{eqnarray}
W^{(u)}_{\rm B 1}&=&\frac {y_u}{\sqrt3} Q^T\left( \begin{matrix} 1&1&1\\1&1&1\\1&1&1\\ \end{matrix}\right)\bar uH_u+y_uQ^T\left(
\begin{array}{ccc}
-\frac{1}{2}- \frac{1}{2\sqrt{3}} &  \frac{1}{\sqrt{3}} & \frac{1}{2}- \frac{1}{2\sqrt{3}}\\  \frac{1}{\sqrt{3}} & \frac{1}{2}- \frac{1}{2\sqrt{3}} & -\frac{1}{2}- \frac{1}{2\sqrt{3}} \\ \frac{1}{2}- \frac{1}{2\sqrt{3}} & -\frac{1}{2}- \frac{1}{2\sqrt{3}} &  \frac{1}{\sqrt{3}}
%
\end{array}
\right)\bar uM_{u1} \nonumber \\
&+& y_u Q^T\left(
\begin{array}{ccc}
\frac{1}{2}- \frac{1}{2\sqrt{3}}&  \frac{1}{\sqrt{3}} & -\frac{1}{2}- \frac{1}{2\sqrt{3}}\\  \frac{1}{\sqrt{3}} & -\frac{1}{2}- \frac{1}{2\sqrt{3}} & \frac{1}{2}- \frac{1}{2\sqrt{3}} \\ -\frac{1}{2}- \frac{1}{2\sqrt{3}} & \frac{1}{2}- \frac{1}{2\sqrt{3}} &  \frac{1}{\sqrt{3}}
%
\end{array}
\right)\bar uM_{u2}.
\end{eqnarray}
The MSSM Yukawa coupling matrix $Y_u$ has one nonzero eigenvalue $\lambda_t=\sqrt{3}$ and two zero eigenvalues.  The nonzero eigenvalue is associated with the eigenvector $(1/\sqrt{3})(1,1,1)$.  The degenerate manifold is spanned by linear combinations of the orthornormal basis set  $(1/\sqrt{2})(-1,1,0)$ and $(-1/\sqrt{6}, -1/\sqrt{6},\sqrt{2/3})$.  Defining the diagonalization matrices $\mathcal{U}_L=\mathcal{U_R}=\mathcal{U}$, such that 
\begin{eqnarray}
\mathcal{U}^\dagger Y_u \mathcal{U} = Y_u^{(\rm diag)}= {\rm Diag} (0,0, \sqrt{3}),
\end{eqnarray}
the matrix $\mathcal{U}=\mathcal{U}(\alpha)$ takes the general form
\begin{eqnarray}
\mathcal{U(\alpha)}=\left (\begin{array}{ccc} -\frac{\cos\alpha}{\sqrt{2}}-\frac{\sin\alpha}{\sqrt{6}} & -\frac{\cos\alpha}{\sqrt{6}}+\frac{\sin\alpha}{\sqrt{2}} & \frac{1}{\sqrt{3}}\\  \frac{\cos\alpha}{\sqrt{2}}-\frac{\sin\alpha}{\sqrt{6}} &  -\frac{\cos\alpha}{\sqrt{6}}-\frac{\sin\alpha}{\sqrt{2}} &  \frac{1}{\sqrt{3}}\\ 
\sqrt{\frac{2}{3}}\sin\alpha & \sqrt{\frac{2}{3}}\cos\alpha & \frac{1}{\sqrt{3}} \end{array} \right ),
\label{umat}
\end{eqnarray}
in which $\alpha$ is a continuous parameter that is included for completeness (it will drop out of all physical observables).  
In the basis that the quarks are diagonal, we have
\begin{eqnarray}
W^{(u)}_{\rm B1}&=& y_u Q_m^T\left( \begin{matrix} 0&0&0\\0&0&0\\0&0&\sqrt{3} \end{matrix}\right)\bar u_mH_u\nonumber \\&+&y_uQ_m^T\left(
\begin{array}{ccc}
-\frac{\sqrt{3}}{2}(\cos 2\alpha+\sin 2\alpha) & \frac{\sqrt{3}}{2}(-\cos 2\alpha+\sin 2\alpha)&0 \\
 \frac{\sqrt{3}}{2}(-\cos 2\alpha+\sin 2\alpha) & \frac{\sqrt{3}}{2}(\cos 2\alpha+\sin 2\alpha) &0 \\
  0 & 0 & 0 \\
\end{array}
\right)\bar u_mM_{u1} \nonumber \\ &+&
y_u Q_m^T\left(
\begin{array}{ccc}
\frac{\sqrt{3}}{2}(-\cos 2\alpha+\sin 2\alpha) & \frac{\sqrt{3}}{2}(\cos 2\alpha+\sin 2\alpha)&0 \\
 \frac{\sqrt{3}}{2}(\cos 2\alpha+\sin 2\alpha) & \frac{\sqrt{3}}{2}(\cos 2\alpha-\sin 2\alpha) &0 \\
0 & 0 & 0 \\
\end{array}
\right)\bar u_mM_{u2}. \label{allequalcouplings}
\end{eqnarray}
We see from Eq.~(\ref{allequalcouplings}) that the top quark does not couple to the messengers, {\it i.e.}, $(Y^\prime_{u1})_{33}=(Y^\prime_{u2})_{33}=0$, in stark contrast from the models of the previous subsection.  The fact that the field with a nonvanishing observable sector Yukawa coupling has a vanishing messenger Yukawa coupling (and vice versa) is a consquence of the $\mathcal{S}_3$ family symmetry (and the enhanced symmetry of taking equal values for the $\beta_i$); this feature was also found in PRZ \cite{Perez:2012mj}.

For this scenario, the soft term contributions from the messenger Yukawas are diagonal in family space, and have nonzero (and degenerate) entries only in the first two generations, as follows:
\begin{eqnarray}
\delta m^2_{Q_{11}}&=& \delta m^2_{Q_{22}}=\frac{\Lambda^2}{(4\pi)^4} \left [  6 Y_t^4+ 6 Y_b^4+ 2 Y_b^2 Y_t^2+  Y_b^2 Y_\tau^2 - \tilde{g}_u^2 Y_t^2- \tilde{g}_d^2 Y_b^2 \right ] \nonumber \\
\delta m^2_{\bar{u}_{11}}&=& \delta m^2_{\bar{u}_{22}}= \frac{\Lambda^2}{(4\pi)^4} \left [  12 Y_t^4+2 Y_t^2 Y_b^2-2 \tilde{g}_u^2 Y_t^2 \right ] \nonumber \\
\delta m^2_{\bar{d}_{11}}&=& \delta m^2_{\bar{d}_{22}}= \frac{\Lambda^2}{(4\pi)^4} \left [  12 Y_b^4+2 Y_t^2 Y_b^2+2 Y_b^2 Y_\tau^2 -2  \tilde{g}_d^2 Y_b^2 \right ] \nonumber \\
\delta m^2_{L_{11}}&=& m^2_{L_{22}}= \frac{\Lambda^2}{(4\pi)^4} \left [  4 Y_\tau^4+ 3 Y_b^2 Y_\tau^2 - \tilde{g}_e^2 Y_\tau^2 \right ] \nonumber \\
\delta m^2_{\bar{e}_{11}}&=& \delta m^2_{\bar{e}_{22}}= \frac{\Lambda^2}{(4\pi)^4} \left [  8 Y_\tau^4+ 6 Y_b^2 Y_\tau^2-2 \tilde{g}_e^2 Y_\tau^2 \right ]\nonumber \\
\delta m^2_{H_u} &=& \delta m^2_{H_d} =0,  \qquad \tilde{A}_u = \tilde{A}_d=\tilde{A}_e =0.
\label{softtermsB1}
\end{eqnarray}
We note that the parameter $\alpha$  of Eq.~(\ref{allequalcouplings}) drops out of the messenger contributions (as it should).

As expected, we see from Eq.~(\ref{softtermsB1}) that the corrections influence only the first two generations, as it is these fields that couple to the messengers.  These corrections are also very likely to be negative for the squarks, pushing the first and second generation below the third generation.  We note in particular that the trilinear couplings vanish for all generations.  For the first and second generations, this is due to the absence of MSSM Yukawa couplings, while for the third generation, it is due to the absence of messenger Yukawas. 

Clearly, this scenario represents only one possible first step toward any kind of realistic theory of this type.  Different ways to perturb the all-equal coupling constraint will result in different predictions for the Yukawa couplings of the lighter generations, which in turn will have correlated predictions for the messenger Yukawas and the soft supersymmetry breaking terms.  We defer a systematic study of these possibilities to future work.

\section{Results}

We now turn to a phenomenological analysis of each of these scenarios.  The parameters shared by all three models include the messenger scale, which is given by the mass of the two heavy doublets, $M_{\rm mess}=M_{2,3}\simeq \sqrt{3/2} M$, and the scale  $\Lambda=F_{2,3}/M_{2,3}\simeq F/M$. The quantities $\mu$ and $b$ are replaced as usual by $\tan\beta$, $\text{sign}(\mu)$, and the $Z$ boson mass, since we are free to tune them independently. 
While the parameters associated with the messenger triplet sector,  $M_T= \langle X_T \rangle$ and $\Lambda_T\equiv F_T/M_T$, are in principle unrelated to $M_{2,3}$ and $\Lambda$, we set them equal for simplicity.   Hence, in these scenarios there are the usual gauge-mediated terms at the messenger scale $M_{2,3}$, and the additional contributions in each case due to the messenger Yukawa couplings.  The parameter $N$, labeling the number of messengers, is always given by $N=2$, and we will always choose $\text{sign}(\mu)$=1.  The renormalization group equations are run using SoftSUSY 4.1.0 \cite{Allanach:2001kg}. 

\subsection{Models A1 and A2: Top quark Yukawa coupling from $\mathcal{H}_u^{(1)}$}

We begin with an analysis of the expected mass spectra in Models A1 and A2, in which the top quark Yukawa coupling arises from a renormalizable coupling to the $\mathcal{S}_3$ singlet Higgs-messenger field $\mathcal{H}_u^{(1)}$.  In both cases, we see that the corrections to the soft scalar mass-squared parameters as given in Eqs.~(\ref{softtermsA1}) and (\ref{softtermsA2}) for Models A1 and A2, respectively, have large contributions from the messenger couplings, especially the top quark messenger couplings.  As a result, in both cases, the gauge part of the corrections will be overwhelmed, leading to positive deflections for the third generation soft mass-squared parameters, and particularly for the stops.  The other large correction occurs in the up-type Higgs soft mass-squared parameter, which is typically large and negative.  
\begin{figure}[htbp] 
   \centering
   \includegraphics[width=3in]{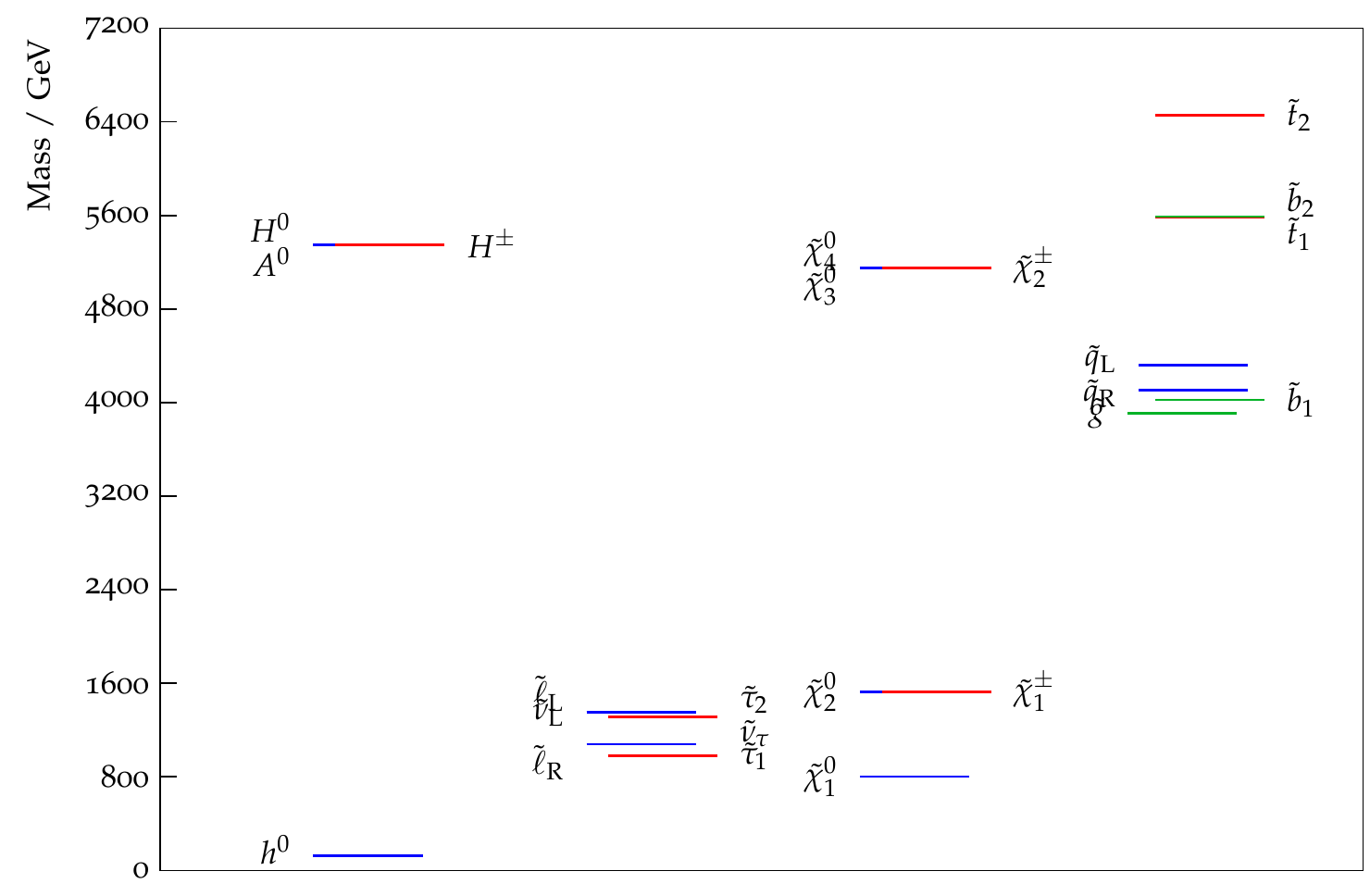} 
    \includegraphics[width=3in]{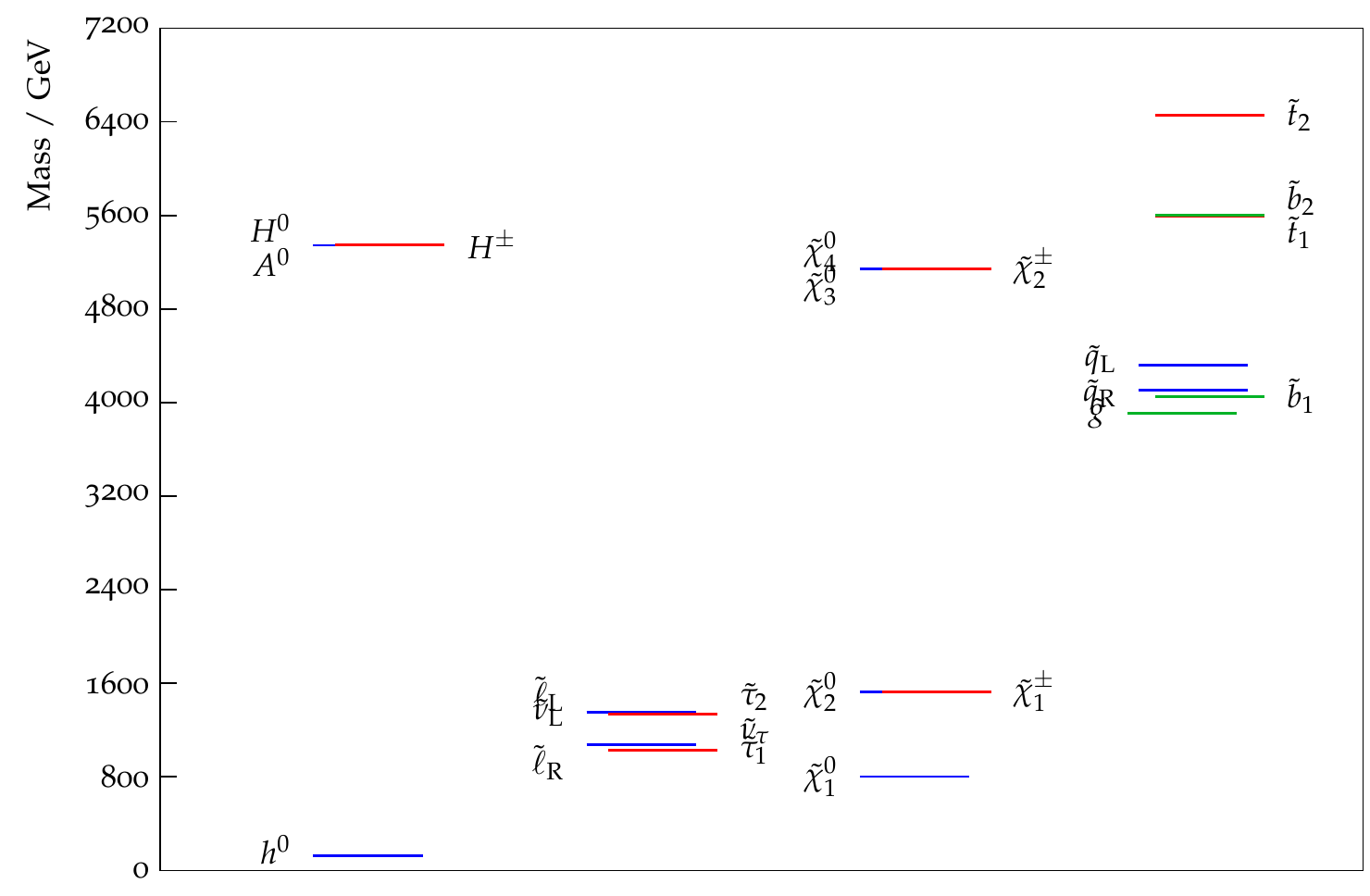} 
   \caption{A comparison of the mass spectra for Models A1 (left) and A2 (right) with a low messenger scale of $M_{\rm mess}=10^6$ GeV,  $\Lambda=2.9\times 10^5$ GeV, and $\tan\beta=10$.}
   \label{fig:A1A2spectralow}
\end{figure}
\begin{figure}[htbp] 
   \centering
   \includegraphics[width=3in]{A1LowScale.pdf} 
    \includegraphics[width=3in]{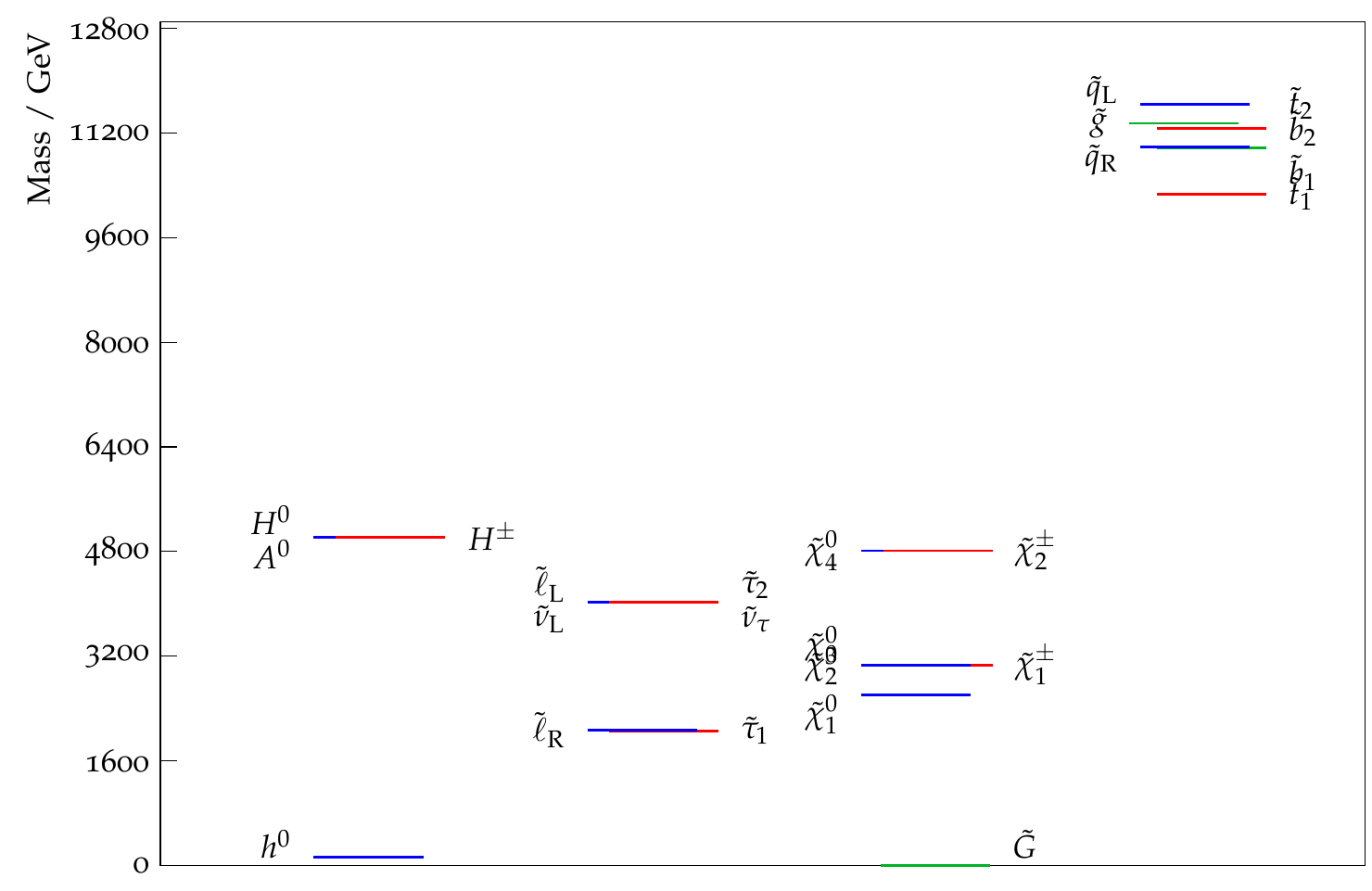} 
   \caption{A comparison of the mass spectra for Model A1(left, identical to the left panel of Fig.~\ref{fig:A1A2spectralow}), and mGMSB with $N=2$, $\Lambda=8.1\times 10^5$ GeV, $M_{\rm mess}=10^6$ GeV, and $\tan\beta=10$ (right).}
   \label{fig:A1mGMSBspectralow}
\end{figure}
\begin{figure}[htbp] 
   \centering
   \includegraphics[width=3in]{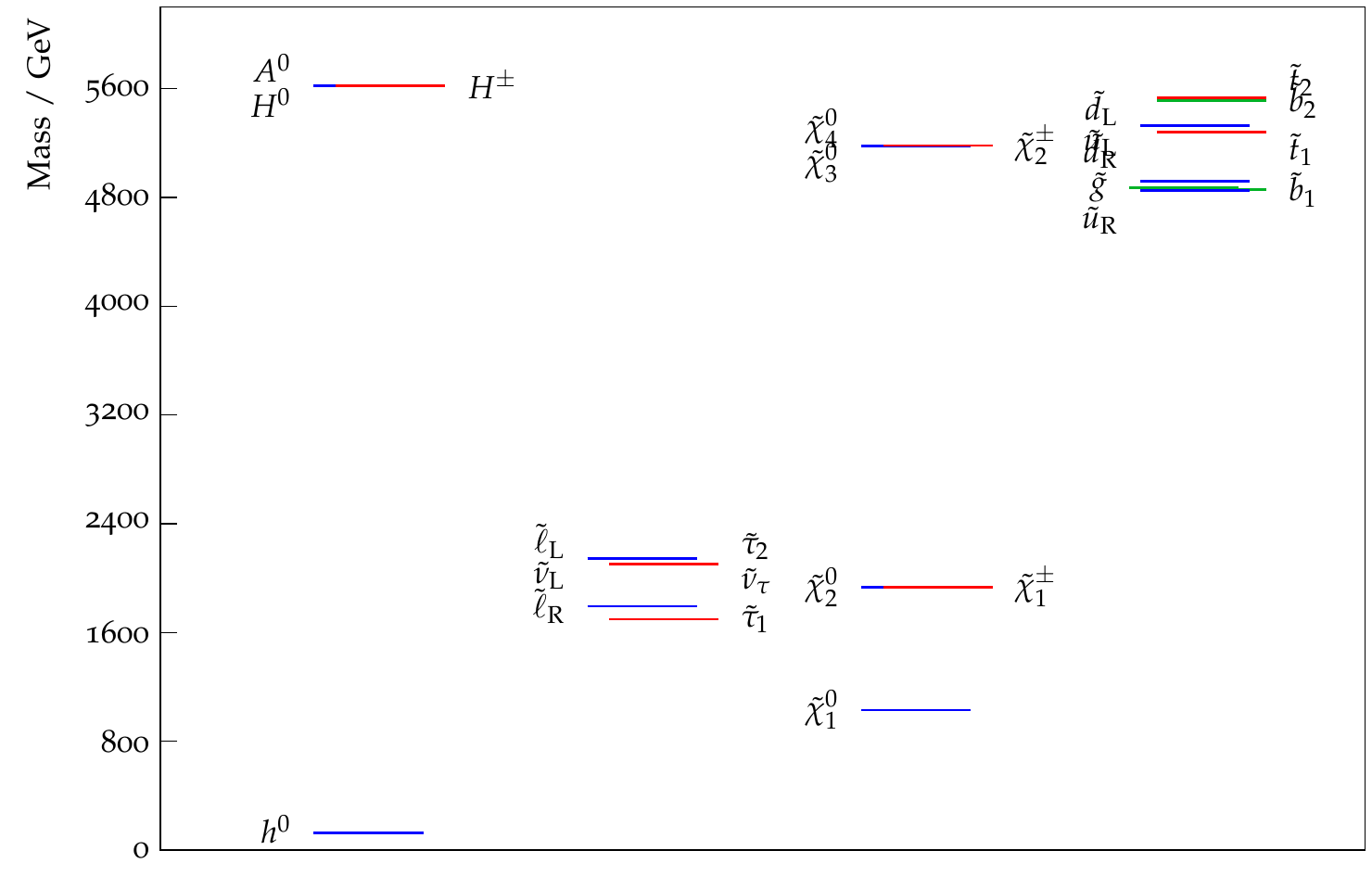} 
    \includegraphics[width=3in]{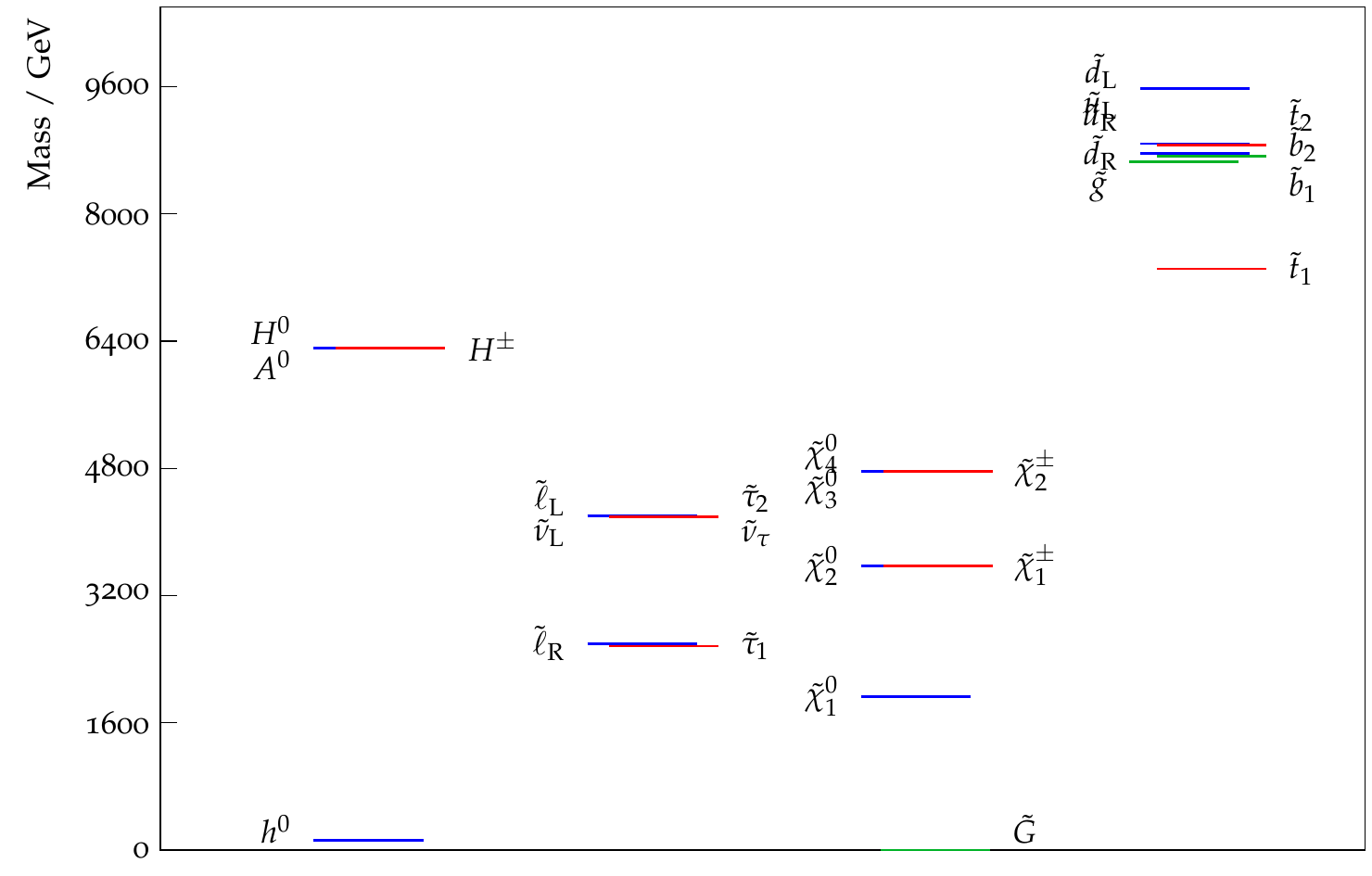} 
   \caption{Model A1(left), with $M_{\rm mess}= 10^{12}$ GeV, $\Lambda = 3.8 \times 10^{5}$ GeV, and $\tan\beta=10$; and mGMSB (right) with $N=2$, $\Lambda=7.0 \times 10^5$ GeV, $M_{\rm mess}=10^{12}$ GeV, and $\tan\beta=10$.}
   \label{fig:A1mGMSBspectraint}
\end{figure}

These features can easily be seen from direct comparisons of Model A1 and Model A2.  In  Fig.~\ref{fig:A1A2spectralow}, we show representative mass spectra for each model for the case of $\tan\beta=10$, $\Lambda = 2.9 \times 10^5$ GeV, and a low messenger scale of  $M_{\rm mess}= 10^6$ GeV.   We see that the two models are highly similar, with only slight differences among the splittings of the squarks and sleptons.  In both cases, the NLSP is the lightest neutralino (which is bino-like), and the lightest colored superpartner is the gluino, with a mass of $m_{\tilde{g}}=3.9$ TeV.  It is notable that the lightest squark is the sbottom $\tilde{b}_1$, which at $m_{\tilde{b}_1}=4.0$ TeV is quite close in mass to the gluino.  The sbottoms are strongly mixed (more so than the stops), with $\tilde{b}_1$ significantly lighter than $\tilde{t}_1$, which has a mass of $m_{\tilde{t}_1}=5.5$ TeV. This behavior arises because of the large messenger couplings to the top quark superfield in both constructions.  These large and positive contributions boost the values of the stop mass-squared parameters  such that the mixing is not as prominent as it is in the sbottom sector.  

It is illuminating to compare these two nearly identical scenarios with minimal gauge mediation models with $N=2$ that can reproduce the observed Higgs mass value of $m_h=125$ GeV.  As is well known, the absence of one-loop contributions to the soft trilinear scalar couplings of the squarks in mGMSB puts strong constraints on the squark mass spectra, particularly for low values of the messenger scale, where there is generically an insufficient amount of renormalization group evolution to yield an appreciable values of the soft trilinear scalar couplings.  This is illustrated in Fig.~\ref{fig:A1mGMSBspectralow}, for which the left panel shows the Model A1 low scale point also presented in the left panel of Fig.~\ref{fig:A1A2spectralow}, and the right panel shows a low messenger scale mGMSB point with $N=2$, $\Lambda=8.1\times 10^5$ GeV, $M_{\rm mess}=10^6$ GeV, and $\tan\beta=10$.  Clearly the spectra are significantly different in the two cases, as expected.  The mGMSB scenario is characterized by  ultraheavy ($\sim 10$ TeV) squarks with the lighter stop as the lightest colored superpartner, and a much larger splitting between the $SU(3)_c$-charged sector and other superpartners than what occurs in Models A1 and A2.

\begin{figure}[htbp] 
   \centering
   \includegraphics[width=3in]{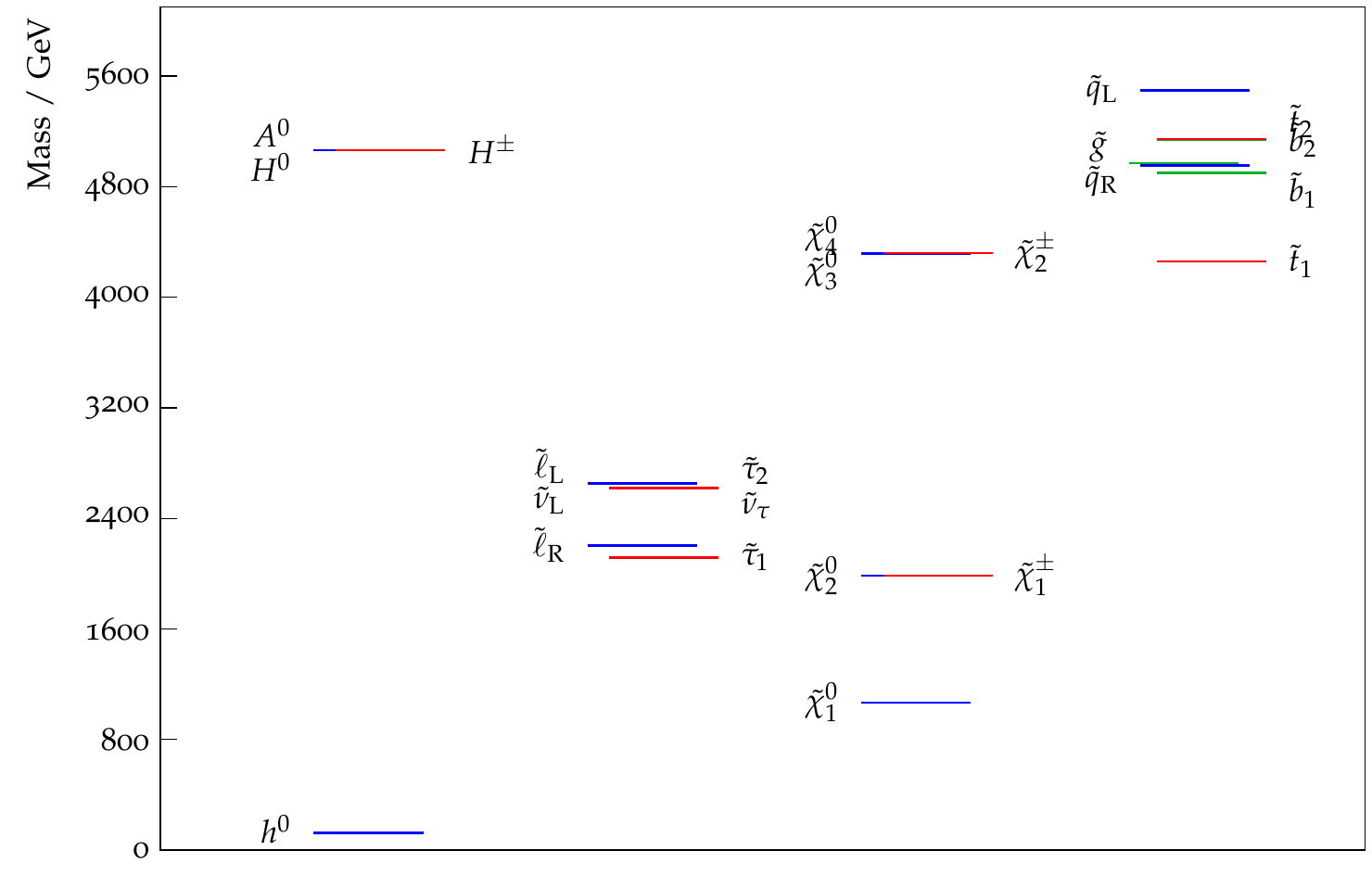} 
    \includegraphics[width=3in]{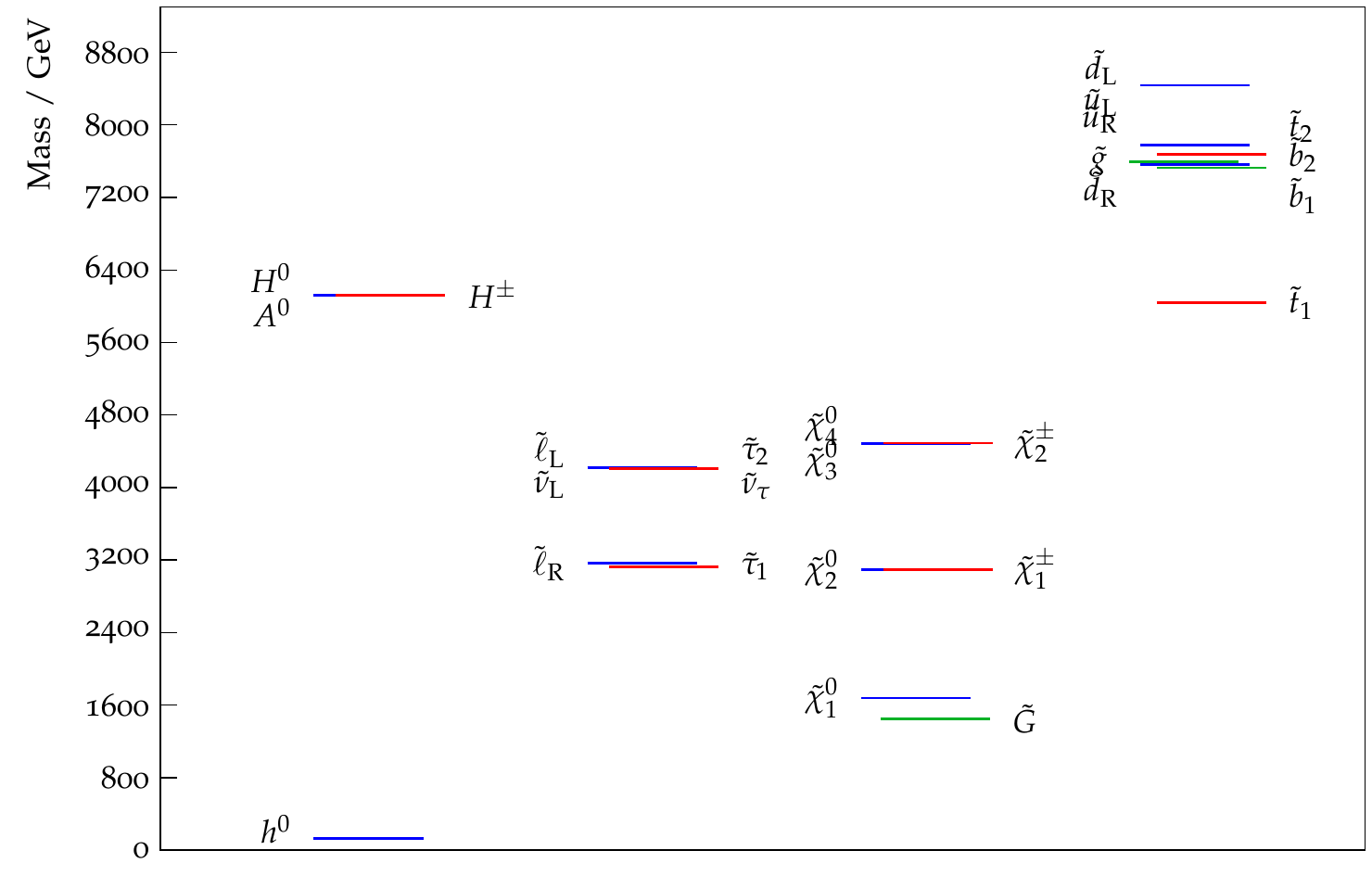} 
   \caption{Model A1(left), with $M_{\rm mess}= 10^{16}$ GeV, $\Lambda = 3.25 \times 10^{5}$ GeV, and $\tan\beta=10$; and mGMSB (right) with $N=2$, $\Lambda=5.7 \times 10^5$ GeV, $M_{\rm mess}=10^{16}$ GeV, and $\tan\beta=10$.}
   \label{fig:A1mGMSBspectrahigh}
\end{figure}
\begin{figure}[htbp] 
   \centering
   \includegraphics[width=4in]{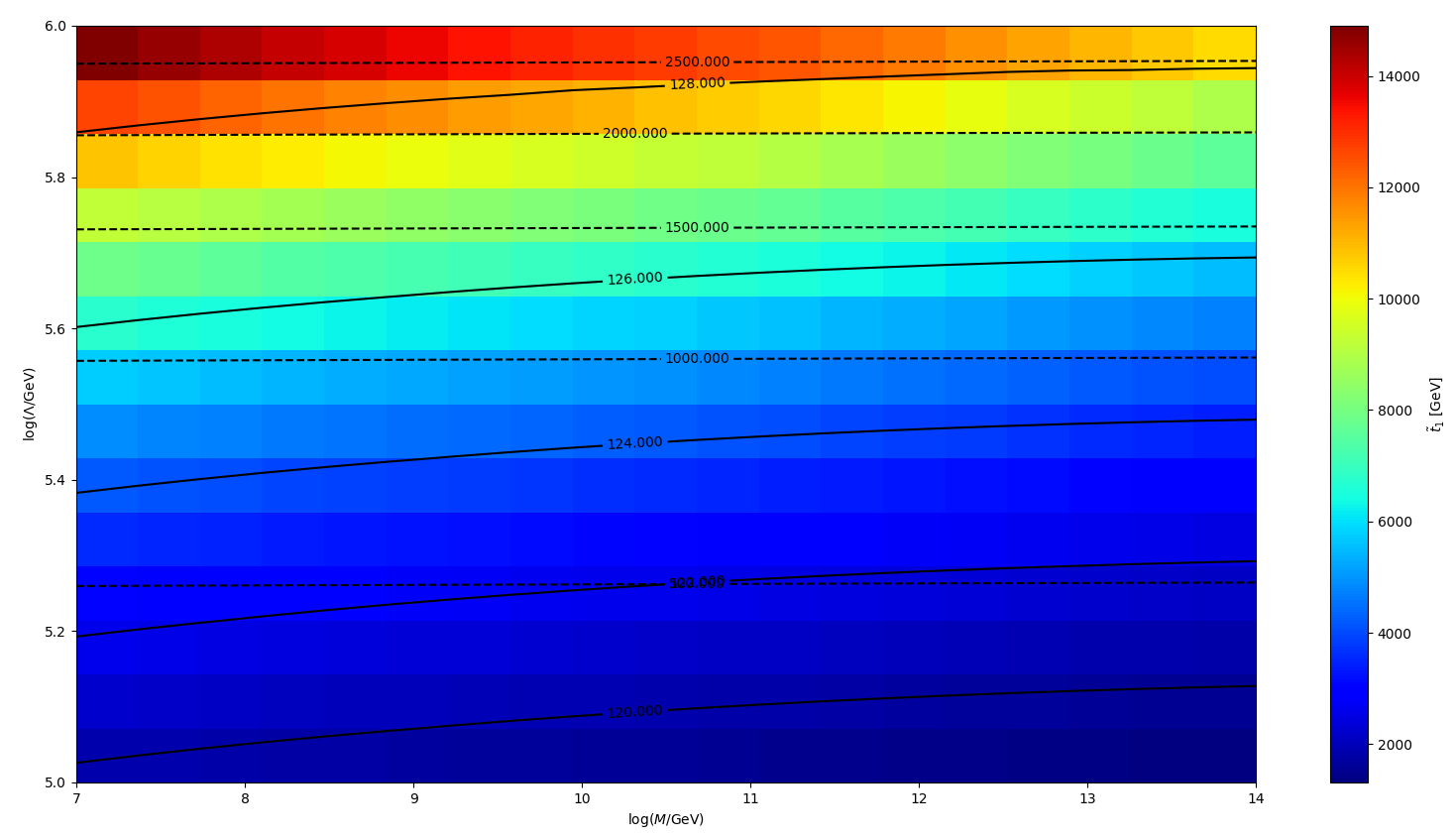} 
   \caption{A scan over the parameter space for Model A1, with $\tan\beta=10$. The solid lines are the Higgs mass, the dotted lines are the bino-like neutralino NLSP, and the color is the stop mass.}
   \label{fig:A1spectrum1}
\end{figure}

These features largely persist for higher values of the messenger scale.  In Fig.~\ref{fig:A1mGMSBspectraint},we show example spectra for Model A1 (Model A2 is roughly identical), with $M_{\rm mess}=10^{12}$ GeV, $\Lambda = 3.8 \times 10^5$ GeV, and $\tan\beta =10$, and minimal GMSB with $N=2$, the same messenger scale of $M_{\rm mess}=10^{12}$ GeV, $\Lambda = 7.0 \times 10^5$ GeV, and $\tan\beta=10$. For both cases, the bottom squarks and the gluino are now heavier, while the lighter stop is lighter, than in the case of low messenger scales.  In Model A1, we have heavier sleptons and charginos/neutralinos than in the case of the low messenger scale, while for mGMSB, the charginos/neutralinos are more strongly split, and the NLSP is now the lightest (bino-like) neutralino, as in the flavored gauge mediation cases.  At values of the messenger scale near the GUT scale, these trends persist for the same value of $\tan\beta$, eventually resulting in the lightest stop as the lightest colored superpartner for Models A1 and A2, as it is in mGMSB.  This behavior is shown in Fig.~\ref{fig:A1mGMSBspectrahigh}.

\begin{figure}[htbp] 
   \centering
   \includegraphics[width=4in]{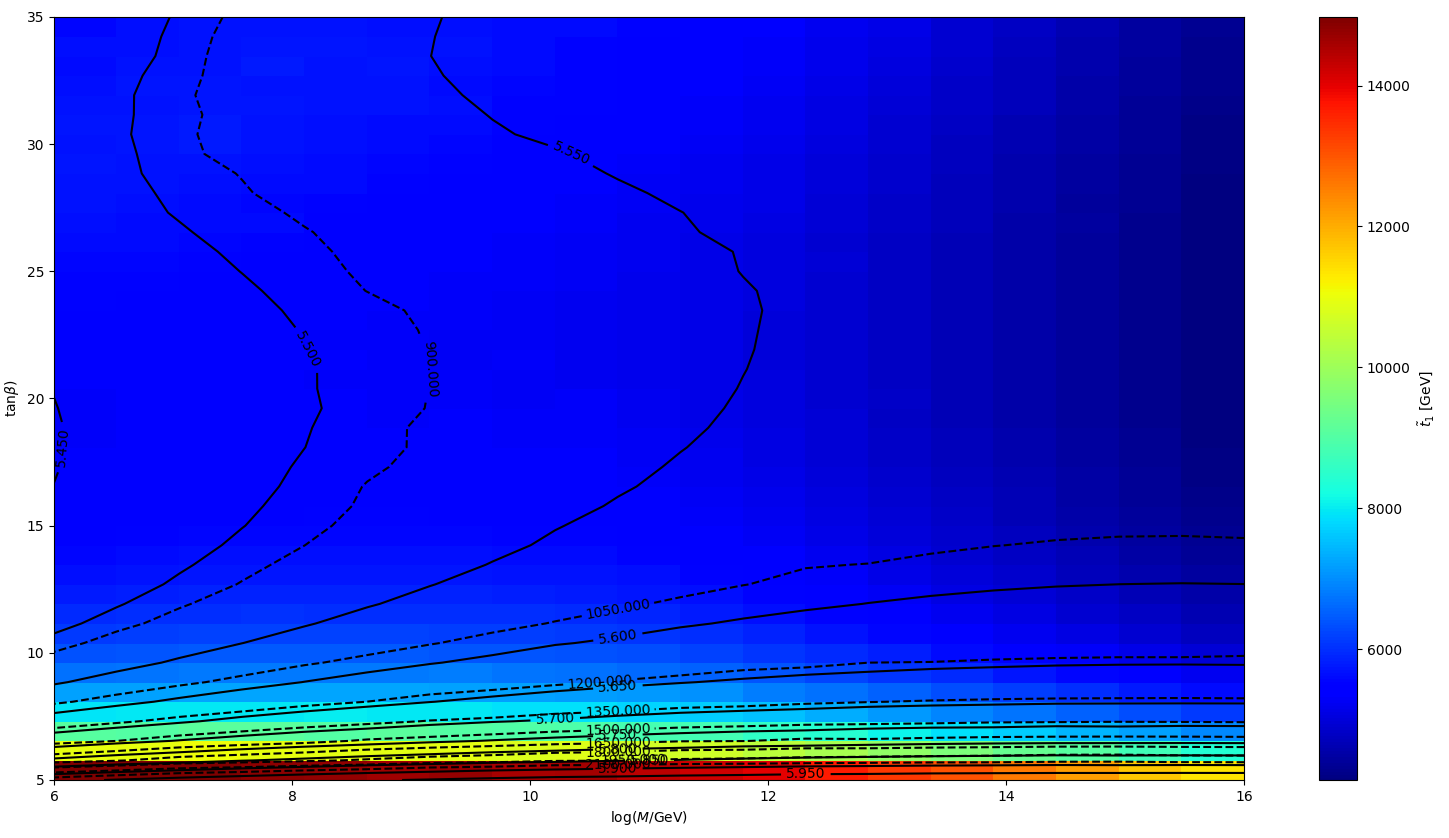} 
   \caption{A scan over the parameter space for Model A1 as a function of $\tan\beta$ and $M_{\rm mess}$, with $\Lambda$ fixed to maintain the light Higgs mass prediction. The solid lines are contours in $\Lambda$, the dotted lines are the lightest neutralino mass, and the color is the stop mass.}
   \label{fig:A1weatherchannel1}
\end{figure}
In Fig.~\ref{fig:A1spectrum1}, we show a scan over the parameter space for Model A1 with $\tan\beta=10$, with solid and dotted contours for the Higgs mass and the lightest neutralino, respectively, and the color denoting the stop mass.  Generically, we would detect a binolike NLSP with nearby right-handed sleptons, and stops in the $4-6$ TeV range, and hence only the left-handed sleptons and the wino would be accessible at the LHC in the near future.    In Fig.~\ref{fig:A1weatherchannel1}, we show the effect of changing $\tan\beta$ and $M_{\rm mess}$ for Model A1, with the value of $\Lambda$ chosen to keep $m_h=125$ GeV. The solid lines are contours in $\Lambda$, the dotted lines are contours in the mass of the lightest neutralino, and the color denotes the lightest stop mass. We see that at low values of $\tan\beta$, stops are typically heavy, because the one-loop correction to the Higgs mass is driven up by the logarithmic term with interference from left-right mixing. This effect is ameliorated for larger values of $\tan\beta$,  allowing for lighter stops. Model A2 displays nearly identical behavior, with nearly exact overlap at low $\tan\beta$, but allowing for slightly smaller values of $\Lambda$ (and hence for the superpartner spectrum) for larger values of $\tan\beta$.

\subsection{Model B1: Top quark Yukawa coupling from $\mathcal{H}_u^{(2)}$}

We now turn to the case of Model B1, in which the top quark Yukawa coupling arises from couplings to $\mathcal{S}_3$ doublet fields.  In this scenario, the MSSM and messenger Yukawas are anticorrelated, with large MSSM Yukawas for the third family fields, and large and diagonal messenger Yukawas only for the first and second generations, as seen in Eq.~(\ref{softtermsB1}).

\begin{figure}[htbp] 
   \centering
   \includegraphics[width=3in]{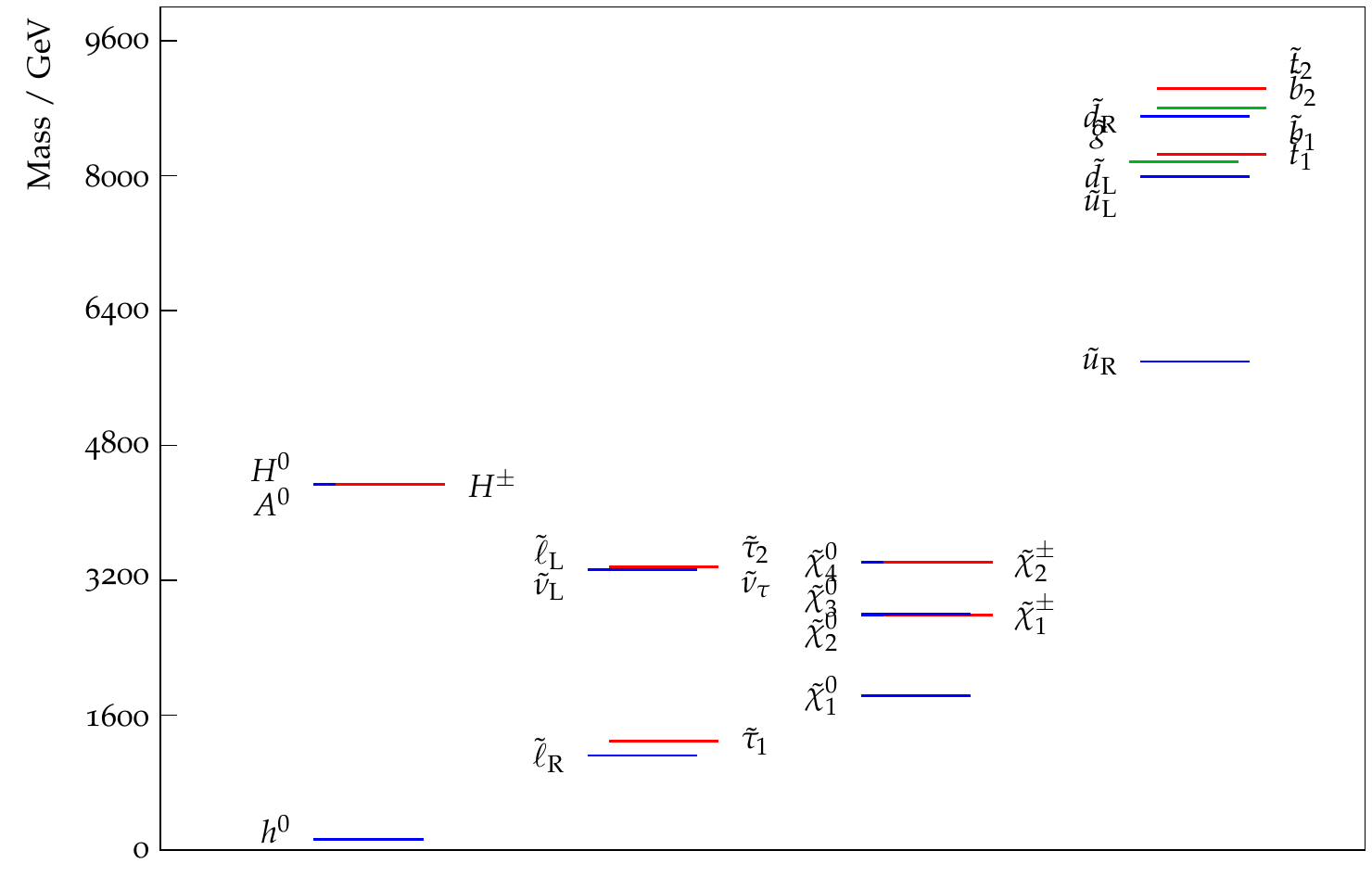} 
   \includegraphics[width=3in]{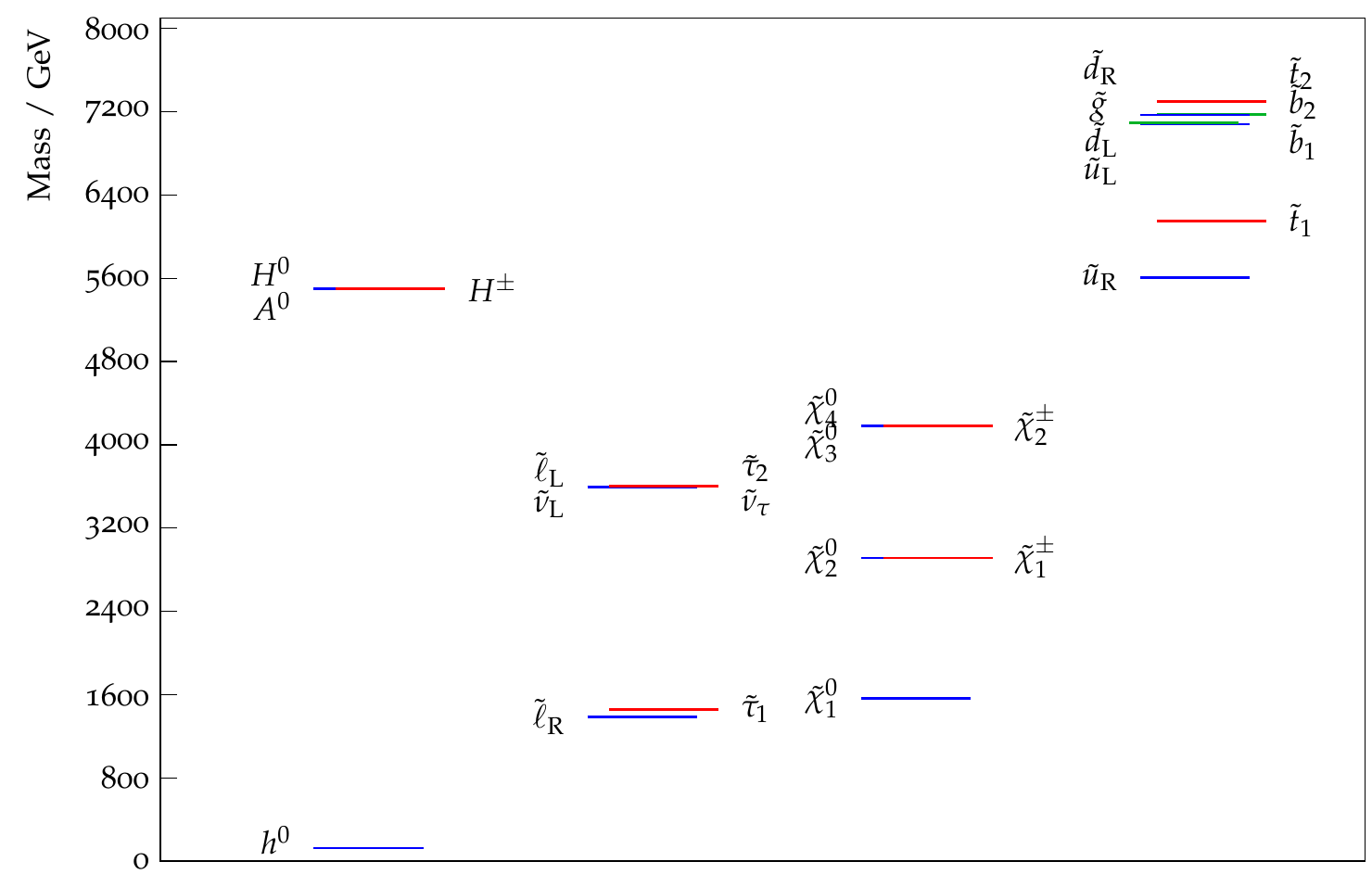} 
   \caption{Two example mass spectra for Model B1, with (i) $M_{\rm mess}=10^6$ GeV, $\Lambda=6.6\times 10^5$ GeV, and $\tan\beta=10$ (left), and (ii) $M_{\rm mess}= 10^{12}$ GeV,  $\Lambda=5.7\times 10^5$ GeV, and $\tan\beta=10$ (right).}
   \label{fig:B1LowInt}
\end{figure}
\begin{figure}[htbp]
\begin{center}
\includegraphics[width=5 in]{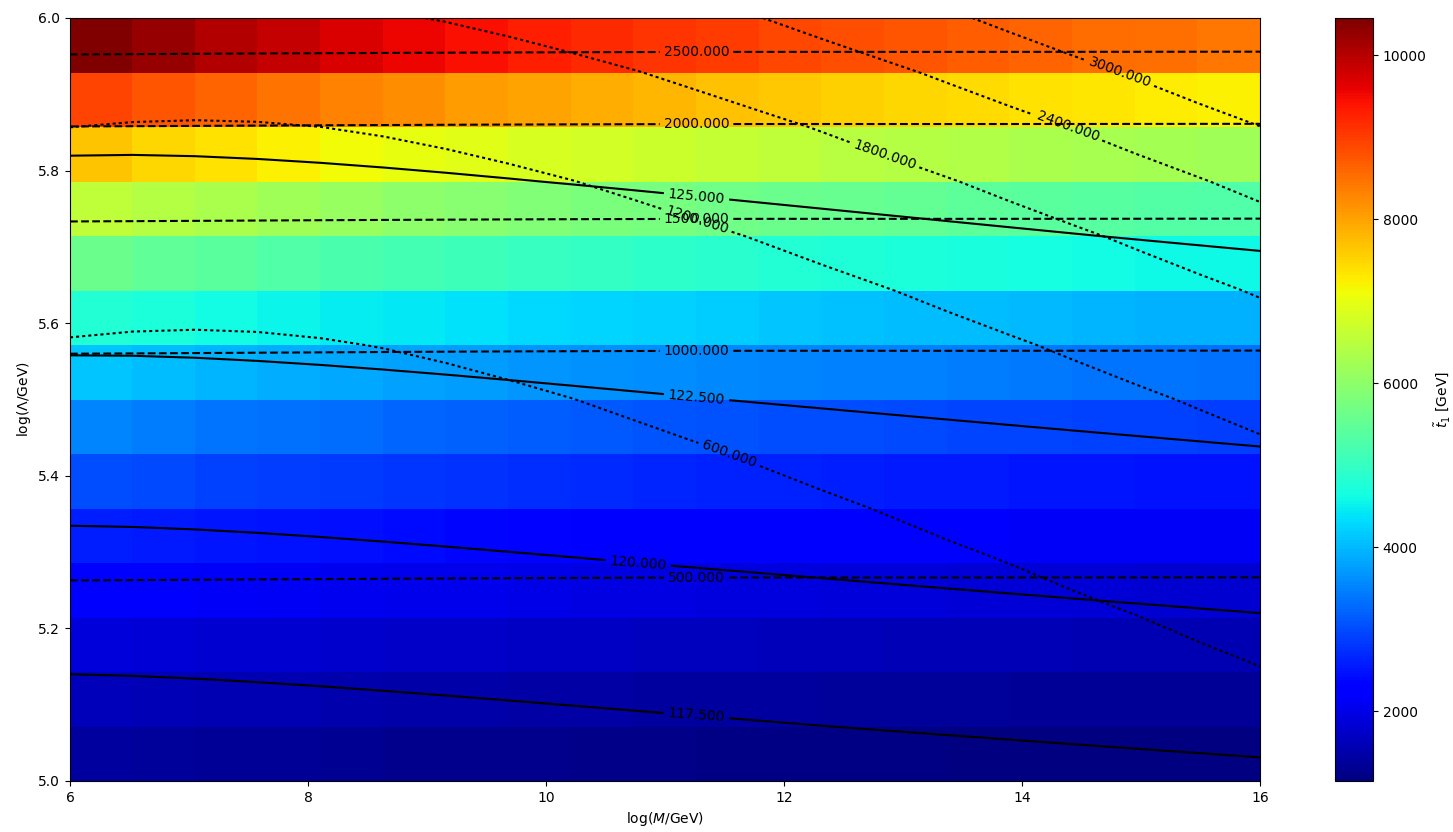}
\caption{The stop mass distribution for Model B1 with $\tan\beta=10$, with contours of the Higgs mass (solid), the lightest slepton mass (dotted), and the lightest neutralino mass (dashed).}
\label{fig:Bspectrum1}
\end{center}
\end{figure}

Two example spectra for Model B1 are shown in Fig.~\ref{fig:B1LowInt}.   In both cases, the Higgs mass in this model is bolstered by heavy stops, because the third generation $A$ terms vanish, resulting in general in heavier spectra than in Models A1 and A2.  In Model B1, the lightest $SU(3)_c$ charged particles are the first and second generation right-handed squarks because the corrections deflect the soft mass-squared parameters down.   In either case, we see that the NLSP will likely be a long-lived right-handed slepton. The corrections from messenger matter mixing generally push the  smuon and selectron below the stau and because the spectra are split, there are large corrections from the running at large $M$ from the $\STr(m^2)$ term in the $\beta$ function. These models can be constrained by searches for charged tracks at LHC13 \cite{Feng:2015wqa}. The lightest neutralino is always binolike and its mass is generally near the masses of the sleptons.

In Fig.~\ref{fig:Bspectrum1}, we show a scan of the parameter space for Model B1, with $\tan\beta=10$. The solid lines are the Higgs mass, the dashed lines are the bino-like neutralino mass, the dotted contours are the lightest slepton mass,  and the color is the stop mass.  One interesting feature is the mass of colored particles at $\sim 7$ TeV, heavier than the case of Models A1 and A2 (as expected).  These models are more strongly constrained as a function of $\tan\beta$, with the possibility of tachyonic sleptons for large $\tan\beta$ and low to intermediate messenger scales once the Higgs mass requirement is imposed.

Here we note that the messenger-matter mixing corrections split the first and second generations from the third generation at the messenger scale, and this difference is ameliorated by renormalization group running towards an IR fixed point. This  ``focusing" behavior was found at low scales in \cite{Perez:2012mj} because they had a larger top quark messenger coupling than what is used here.

\subsection{Discussion}

We have seen that these models allow for viable superpartner spectra while achieving $m_h\sim 125$ GeV.  For the scenarios in which it is the third family that has nonvanishing messenger Yukawa couplings (Models A1 and A2), the spectra for $\tan\beta=10$ are characterized by squarks and gluinos in the $4-6$ TeV range and a bino-like NLSP neutralino.  In the case in which it is the first and second generations that have nontrivial messenger Yukawas (Model B1), the squarks and gluinos are heavier ($\sim 7$ TeV), and the lightest squark is one of the first or second generation squarks. 

It is instructive to compare these scenarios with other representative examples of flavored gauge mediation in the literature, such as the family $U(1)$ benchmark models of \cite{Ierushalmi:2016axs}, all of which involve messenger Yukawa couplings in the up quark sector only.  These benchmark models are of course far more developed than the toy scenarios considered here, as they have a full treatment of the three-family MSSM Yukawa couplings and the three-family messenger Yukawa couplings to the up-type quarks, which allows for reliable estimates of the flavor-mixing effects in the soft supersymmetry-breaking scalar mass-squared parameters.  Nonetheless, the spectra of some of these benchmark examples resemble our models to some extent, for example, with the up squark or charm squark as the lightest $SU(3)_c$-charged superpartner for the non-minimal-flavor-violating cases with a nonvanishing messenger Yukawa coupling only to the first or second generation up-type quarks, as in our Model B1 in which there is a diagonal coupling in the first and second generation subblock.

However, one main difference is that the family $U(1)$ benchmark models allow for the possibility that only one of the messenger fields couples to the up-type quarks, instead of two messenger pairs as in the scenarios considered here.  This feature contributes to the fact that these benchmarks have viable spectra that reproduce the light Higgs mass with significantly lighter $SU(3)_c$ charged superpartners, resulting in an improved discovery potential at the LHC.  In our scenarios, the large messenger Yukawa contributions to the soft scalar mass-squared parameters of the squarks result quite generally in large stop masses. Furthermore, our models all include messenger Yukawas in the slepton sector, which typically results in heavier/more split sleptons.

Another important difference between these models and our scenarios is that while they allow for the dominant one-loop contributions to the soft supersymmetry breaking terms that generically arise in flavored gauge mediation, we explicitly forbid these terms through our requirement that $[ \mathbb{M},\mathbb{F}] =0$.  This feature was important in our scenarios for a smooth decoupling of the heavy messengers from the light electroweak Higgs doublets.  However, since the one-loop contributions are generically negative while the two-loop corrections are typically positive, their inclusion can also be an important factor for achieving mass spectra that are accessible at the LHC.

\section{Conclusions}

In this paper, we have explored flavored gauge mediation models in which the electroweak Higgs doublets and the $SU(2)$ messenger doublets are embedded in representations of a Higgs-messenger discrete symmetry group, which we take for concreteness to be the discrete group $\mathcal{S}_3$.  The idea of connecting the Higgs and messenger doublets with a non-Abelian discrete symmetry was first explored in a two-family context by Perez, Ramond, and Zhang \cite{Perez:2012mj}, in which they went a step further and had the same $\mathcal{S}_3$ group also serve as a family symmetry group.  In these scenarios, the supersymmetry-breaking field is a doublet representation of $\mathcal{S}_3$; the field space directions of its scalar and F-component vacuum expectation values generically must be misaligned in order to produce working models with a smooth decoupling of the light Higgs fields from the heavier messenger fields.  The question was whether the intriguing two-family examples of \cite{Perez:2012mj} could be extended into realistic three-family scenarios.
This paper represents a first step in this direction.

We have shown that a model framework of this type, in which the Higgs and doublet messenger fields are taken to be components of Higgs-messenger fields with $\mathcal{S}_3$ quantum numbers, generically suffers from a severe $\mu/B_\mu$ problem.  The reason is that if there is a coupling of the doublet messengers to the supersymmetry breaking field, as is generally needed to mediate supersymmetry breaking, the $\mathcal{S}_3$ symmetry dictates that there will also necessarily be a dangerous direct coupling of the Higgs fields to the supersymmetry breaking field, which results in an incorrect $\mu/B_\mu$ hierarchy.  This problem can be alleviated by expanding the field degrees of freedom to allow for the possibility of independently tuning $\mu$ and $B_\mu$; we achieve this here by allowing for a larger  Higgs-messenger sector that includes both $\mathcal{S}_3$ doublets and $\mathcal{S}_3$ singlets.  While not a satisfactory solution to the $\mu/B_\mu$ problem in that it involves two fine-tunings  (that each increase for higher values of the messenger scale), it does allow for the construction of viable scenarios in which one pair of Higgs-messenger fields is light and thus is identified as the electroweak Higgs fields, while the others are heavy messenger fields that have nontrivial Yukawa couplings to the MSSM fields.  The addition of the singlet Higgs-messenger fields also allows for new possibilities for obtaining a renormalizable top quark Yukawa coupling, which is also a crucial model-building ingredient.

To this end, we have constructed three model scenarios that include only third-family Yukawa couplings of the MSSM fields to the electroweak Higgs fields.  Two of these models (Models A1 and A2) have the top quark Yukawa coupling arising from the Higgs-messenger singlets and differ only in their treatment of the bottom and tau Yukawa couplings; both scenarios predict third family messenger Yukawa couplings of similar size and strength to the MSSM Yukawa couplings.  In these scenarios, there is a one-loop contribution to the trilinear stop coupling, which allows for a viable prediction of the light Higgs mass without ultraheavy squarks as in minimal gauge mediation.  In the third scenario (Model B1), the MSSM fields are also charged under the $\mathcal{S}_3$ symmetry, such that the top quark Yukawa coupling has a nontrivial contribution from the Higgs-messenger $\mathcal{S}_3$ doublets.  In this minimal scenario in which only the third family fermions obtain nonzero masses, the resulting messenger Yukawa couplings are zero for the third family, but nonzero and diagonal in the first and second generation sector.  This results in vanishing soft trilinear scalar couplings, and hence heavier stops are needed to generate the Higgs mass.  The spectra in all cases have squarks in $4-7$ TeV range due to the large messenger corrections that arise from the effective $N=2$ structure of the messenger sector.  Nonetheless, we find it very encouraging that these toy scenarios, which each have a very small number of parameters, allow for a variety of viable spectra.

As stated, these models represent a first step in this direction.  Obtaining realistic models requires that the MSSM Yukawa couplings are fully modelled.  The question of whether the resulting correlated messenger Yukawa couplings can survive stringent flavor constraints is not clear (though it has been pointed out that the flavor-dependent couplings that generically arise in flavored gauge mediation are not as dangerous as it might naively appear~\cite{Calibbi:2014yha,Ierushalmi:2016axs}).  It is worth noting, however, that this framework provides a new and potentially fruitful playground for flavor model building, depending on whether the Higgs-messenger symmetry group also plays the role of a family symmetry group.   There are also other fundamental questions to be addressed, such as possible connections between the $\mathcal{S}_3$ singlet supersymmetry breaking field $X_T$ that couples to the messenger $SU(3)_c$ triplet fields and the $\mathcal{S}_3$ doublet supersymmetry breaking field $X_H$ (in particular, the question of a supersymmetric CP problem since these degrees of freedom are {\it a priori} independent), and the origin of the needed misalignment between the scalar and F-component vacuum expectation values and its connection to the $\mu/B_\mu$ problem.  Studies along these lines are underway.

Exploring novel model-building directions is important to ensure we are able to understand and interpret the outcome of the current unprecedented exploration of the TeV scale at the LHC.  Flavored gauge mediation models quite generally represent a nontrivial extension of minimal gauge mediation that allows for viable MSSM spectra.  In this version in which the Higgs and messenger doublets are connected by a discrete non-Abelian symmetry, we believe that the first model-building steps taken here show promise that in a more complete implementation, this approach may provide useful input for the comprehensive LHC tests of the paradigm of TeV-scale supersymmetry.

\begin{acknowledgments}

L.L.E. is grateful to I.-W.~Kim, D.~J~H.~Chung, M.~McNanna, L.-T.~Wang, and Y.~Zhao for their helpful input.  This work is supported by the U. S. Department of Energy under the contracts DE-FG-02-95ER40896 and DE-SC0017647.  L.L.E. also thanks the Enrico Fermi Institute of the University of Chicago for their support and hospitality during the earlier stages of this work.
\end{acknowledgments}

 \appendix

\section{Corrections from having two sets of messengers that couple to matter}
\label{flavorcorrections}

Reported below are the corrections to the soft mass parameters when there are multiple Higgs-messenger pairs, following the general analysis of Evans and Shih \cite{Evans:2013kxa}. In the limit in which there is only one pair of messengers, there is a suppressed one-loop contribution as well as the two-loop contributions from messenger-matter mixing  for each soft mass-squared parameter, as reported in~\cite{Abdullah:2012tq}. For completeness, we reproduce the dominant two-loop corrections to the soft mass-squared parameters and the one-loop contributions to the soft trilinear scalar couplings in the case of one messenger pair (here labeled by $i$, no sum over repeated indices): 
\begingroup\makeatletter\def\f@size{9}\check@mathfonts
\begin{equation}
\begin{aligned}
\small
\delta_im^2_{Q}&= \frac{\Lambda^2}{(4 \pi)^4}\bigg[\left(3\text{Tr}\left(Y_{ui}'^\dagger Y_{ui}'\right)-\frac {16} 3 g_3^2-3 g_2^2-\frac {13}{15}g_1^2\right)Y_{ui}'Y_{ui}'^\dagger +3Y_{ui}' Y_{ui}'^\dagger Y_{ui}' Y_{ui}'^\dagger \\
&\ \ \ +\left(\text{Tr}\left(3Y_{di}'^\dagger Y_{di}'+Y_{ei}'^\dagger Y_{ei}'\right)-\frac {16} 3 g_3^2-3 g_2^2-\frac {7}{15}g_1^2\right)Y_{di}'Y_{di}'^\dagger+3Y_{di}' Y_{di}'^\dagger Y_{di}' Y_{di}'^\dagger\\
& \ \ \ +Y_{ui}' Y_{ui}'^\dagger  Y_{di}' Y_{di}'^\dagger +Y_{di}' Y_{di}'^\dagger Y_{ui}' Y_{ui}'^\dagger +2 Y_{ui}'Y_{ui}^\dagger Y_{ui} Y_{ui}'^\dagger +2Y_{di}' Y_{di}^\dagger Y_{di}Y_{ui}'^\dagger  \\
&\ \ \ -2Y_{ui} Y_{ui}'^\dagger Y_{ui}' Y_{ui}^\dagger+3Y_{ui}' Y_{ui}^\dagger \text{Tr}\left(Y_{ui}^\dagger Y_{ui}'\right)+3Y_{ui} Y_{ui}'^\dagger \text{Tr}\left(Y_{ui}'^\dagger Y_{ui}\right)\\
& \ \ \ -2Y_{di}Y_{di}'^\dagger Y_{di}'Y_{di}^\dagger +Y_{di}' Y_{di}^\dagger \text{Tr}\left(3Y_{di}^\dagger Y_{di}'+Y_{ei}^\dagger Y_{ei}'\right)+Y_{di} Y_{di}'^\dagger \text{Tr}\left(3Y_{di}'^\dagger Y_{di}+Y_{ei}'^\dagger Y_{ei}\right)
\bigg],
\end{aligned}
\end{equation}
\begin{equation}
\begin{aligned}
\delta_i m^2_{\bar u}&= \frac{\Lambda^2}{(4 \pi)^4} \bigg[2\left(3\text{Tr}\left(Y_{ui}'^\dagger Y_{ui}'\right)-\frac {16} 3 g_3^2-3 g_2^2-\frac {13}{15}g_1^2\right)Y_{ui}'^\dagger Y_{ui}'+6Y_{ui}'^\dagger Y_{ui}'Y_{ui}'^\dagger Y_{ui}'\\
& \ \ \ +2Y_{ui}'^\dagger Y_{ui}Y_{ui}^\dagger Y_{ui}'+2Y_{ui}'^\dagger Y_{di}Y_{di}^\dagger Y_{ui}'+2Y_{ui}'^\dagger Y_{di}'Y_{di}'^\dagger Y_{ui}'-2Y_{ui}^\dagger Y_{ui}'Y_{ui}'^\dagger Y_{ui}\\
&\ \ \ -2Y_{ui}^\dagger Y_{di}'Y_{di}'^\dagger Y_{ui}+6Y_{ui}'^\dagger Y_{ui}\text{Tr}\left(Y_{ui}^\dagger Y_{ui}'\right)+6Y_{ui}^\dagger Y_{ui}'\text{Tr}\left(Y_{ui}'^\dagger Y_{ui}\right)\bigg],
\end{aligned}
\end{equation}
\begin{equation}
\begin{aligned}
\delta_i m^2_{\bar d}&= \frac{\Lambda^2}{(4 \pi)^4} \bigg[2\left(\text{Tr}\left(3Y_{di}'^\dagger Y_{di}'+Y_{ei}'^\dagger Y_{ei}'\right)-\frac {16} 3 g_3^2-3 g_2^2-\frac {7}{15}g_1^2\right)Y_{di}'^\dagger Y_{di}'+6Y_{di}'^\dagger Y_{di}'Y_{di}'^\dagger Y_{di}'\\
& \ \ \ +2Y_{di}'^\dagger Y_{di}Y_{di}^\dagger Y_{di}'+2Y_{di}'^\dagger Y_{ui}Y_{ui}^\dagger Y_{di}'+2Y_{di}'^\dagger Y_{ui}'Y_{ui}'^\dagger Y_{di}'-2Y_{di}^\dagger Y_{di}'Y_{di}'^\dagger Y_{di}-2Y_{dI}^\dagger Y_{ui}'Y_{ui}'^\dagger Y_{di}\\
&\ \ \ +2Y_{di}'^\dagger Y_{di}\text{Tr}\left(3Y_{di}^\dagger Y_{di}'+Y_{ei}^\dagger Y_{ei}'\right)+2Y_{di}^\dagger Y_{di}'\text{Tr}\left(3Y_{di}'^\dagger Y_{di}+Y_{ei}'^\dagger Y_{ei}\right)
\bigg],
\end{aligned}
\end{equation}
\begin{equation}
\begin{aligned}
\delta_i m^2_{L}&= \frac{\Lambda^2}{(4 \pi)^4} \bigg[\left(\text{Tr}\left(3Y_{di}'^\dagger Y_{di}'+Y_{ei}'^\dagger Y_{ei}'\right)-3 g_2^2-\frac {9}{5}g_1^2\right)Y_{ei}' Y_{ei}'^\dagger\\
& \ \ \ +3Y_{ei}'Y_{ei}'^\dagger Y_{ei}' Y_{ei}'^\dagger+2Y_{ei}' Y_{ei}^\dagger Y_{ei} Y_{ei}'^\dagger-2Y_{ei} Y_{ei}'^\dagger Y_{ei}' Y_{ei}^\dagger\\
&\ \ \ +Y_{ei}' Y_{ei}^\dagger\text{Tr}\left(3Y_{di}^\dagger Y_{di}'+Y_{ei}^\dagger Y_{ei}'\right)+Y_{ei} Y_{ei}'^\dagger\text{Tr}\left(3Y_{di}'^\dagger Y_{di}+Y_{ei}'^\dagger Y_{ei}\right)
\bigg],
\end{aligned}
\end{equation}
\begin{equation}
\begin{aligned}
\delta_i m^2_{\bar e}&=\frac{\Lambda^2}{(4 \pi)^4}\bigg[2\left(\text{Tr}\left(3Y_{di}'^\dagger Y_{di}'+Y_{ei}'^\dagger Y_{ei}'\right)-3 g_2^2-\frac {9}{5}g_1^2\right)Y_{ei}'^\dagger Y_{ei}'\\
& \ \ \ +6Y_{ei}'^\dagger Y_{ei}'Y_{ei}'^\dagger Y_{ei}' +2Y_{ei}'^\dagger Y_{ei} Y_{ei}^\dagger Y_{ei}'-2Y_{ei}^\dagger Y_{ei}'Y_{ei}'^\dagger Y_{ei}\\
&\ \ \ +2Y_{ei}'^\dagger Y_{ei}\text{Tr}\left(3Y_{di}^\dagger Y_{di}'+Y_{ei}^\dagger Y_{ei}'\right)+2Y_{ei}^\dagger Y_{ei}'\text{Tr}\left(3Y_{di}'^\dagger Y_{di}+Y_{ei}'^\dagger Y_{ei}\right)
\bigg],
\end{aligned}
\end{equation}
\begin{equation}
\begin{aligned}
\delta_i m^2_{H_u}&=\frac{\Lambda^2}{(4 \pi)^4}\bigg[-3\text{Tr}\left(Y_{ui}^\dagger Y_{ui}' Y_{ui}'^\dagger Y_{ui}+Y_{ui}^\dagger Y_{di}' Y_{di}'^\dagger Y_{ui} +2Y_{ui}^\dagger Y_{ui} Y_{ui}'^\dagger Y_{ui}'\right)
\bigg],
\end{aligned}
\end{equation}
\begin{equation}
\begin{aligned}
\delta_i m^2_{H_d}&=\frac{\Lambda^2}{(4 \pi)^4}\bigg[-3\text{Tr}\left(Y_{di}^\dagger Y_{ui}' Y_{ui}'^\dagger Y_{di}+Y_{di}^\dagger Y_{di}' Y_{di}'^\dagger Y_{di} +2Y_{di}^\dagger Y_{di} Y_{di}'^\dagger Y_{di}'\right)\\
&\ \ \ -3\text{Tr}\left(Y_{ei}^\dagger Y_{ei}' Y_{ei}'^\dagger Y_{ei} +2Y_{ei}^\dagger Y_{ei} Y_{ei}'^\dagger Y_{ei}'\right)
\bigg],
\end{aligned}
\end{equation}
\endgroup
\begin{equation}
\begin{aligned}
\tilde{A}_{ui}^*&=-\frac {\Lambda}{(4\pi)^2}\left(\left(Y_{ui}'Y_{ui}'^\dagger+Y_{di}'Y_{di}'^\dagger \right)Y_{ui}+2Y_{ui}Y_{ui}'^\dagger Y_{ui}'\right),\\
\tilde{A}_{di}^*&=-\frac {\Lambda}{(4\pi)^2}\left(\left(Y_{ui}'Y_{ui}'^\dagger+Y_{di}'Y_{di}'^\dagger \right)Y_{di}+2Y_{di}Y_{di}'^\dagger Y_{di}'\right),\\
\tilde{A}_{ei}^*&=-\frac {\Lambda}{(4\pi)^2}\left(Y_{ei}'Y_{ei}'^\dagger Y_{ei}+2Y_{ei} Y_{ei}'^ \dagger Y_{ei}'\right).
\end{aligned}
\end{equation}
In the situation with more than one pair of messengers (in our case Higgs-messenger pairs), in addition to summing over the index $i$ to include all pairs, there are corrections to the soft mass-squared parameters of the MSSM matter fields from couplings between the pairs of messengers.  Hence, for each sfermion field $f$, we have
\begin{eqnarray}
\delta m^2_f = \sum_i \delta_i m^2_f+\sum_{i>j} \sum_j \delta_{ij} m^2_f,
\end{eqnarray}
in self-evident notation.  For the case of interest here, in which there are effectively two messenger pairs upon diagonalizing the Higgs-messenger sector, the corrections $\delta_{12}m^2_f$ are given by:
\begingroup\makeatletter\def\f@size{9}\check@mathfonts
\begin{equation}
\begin{aligned}
\delta_{12}m^2_{Q}&=\frac{\Lambda^2}{(4 \pi)^4}\bigg[3Y_{u1}' Y_{u2}'^\dagger\Tr\left(Y_{u2}'Y_{u1}'^\dagger\right)+3Y_{u2}' Y_{u1}'^\dagger  \Tr\left(Y_{u1}'Y_{u2}'^\dagger \right)+2\left(Y_{u1}' Y_{u1}'^\dagger Y_{u2}' Y_{u2}'^\dagger +Y_{u2}' Y_{u2}'^\dagger  Y_{u1}'Y_{u1}'^\dagger \right)\\
&\ \ \ \ \ \left(Y_{u1}' Y_{u2}'^\dagger Y_{u2}' Y_{u1}'^\dagger +Y_{u2}' Y_{u1}'^\dagger Y_{u1}' Y_{u2}'^\dagger \right)+\left(Y_{u1}' Y_{d2}'^\dagger  Y_{d2}' Y_{u1}'^\dagger +Y_{u2}' Y_{d1}'^\dagger Y_{d1}' Y_{u2}'^\dagger \right)\\
& \ \ \ \ \ + Y_{d1}'Y_{d2}'^\dagger \Tr\left(3Y_{d2}' Y_{d1}'^\dagger+Y_{e2}' Y_{e1}'^\dagger\right)+Y_{d2}'Y_{d1}'^\dagger  \Tr\left(3Y_{d1}' Y_{d2}'^\dagger +Y_{e1}'Y_{e2}'^\dagger \right)\\
&\ \ \ \ \ +2\left(Y_{d1}' Y_{d1}'^\dagger Y_{d2}' Y_{d2}'^\dagger +Y_{d2}'Y_{d2}'^\dagger Y_{d1}'Y_{d1}'^\dagger \right)+\left(Y_{d1}'Y_{d2}'^\dagger  Y_{d2}'Y_{d1}'^\dagger +Y_{d2}' Y_{d1}'^\dagger Y_{d1}' Y_{d2}'^\dagger \right)\\
& \ \ \ \ \ +\left(Y_{d1}'Y_{u2}'^\dagger  Y_{u2}' Y_{d1}'^\dagger +Y_{d2}' Y_{u1}'^\dagger Y_{u1}' Y_{d2}'^\dagger \right)\bigg],
\end{aligned}
\end{equation}
\begin{equation}
\begin{aligned}
\delta_{12}m^2_{\bar u}&=\frac{\Lambda^2}{(4 \pi)^4}\bigg[6Y_{u1}'^\dagger Y_{u2}'\Tr\left(Y_{u2}'Y_{u1}'^\dagger\right)+6Y_{u2}'^\dagger Y_{u1}' \Tr\left(Y_{u1}'Y_{u2}'^\dagger \right)+4\left(Y_{u1}'^\dagger Y_{u1}'Y_{u2}'^\dagger Y_{u2}'+Y_{u2}'^\dagger Y_{u2}' Y_{u1}'^\dagger Y_{u1}'\right)\\
&\ \ \ \ \ 2\left(Y_{u1}'^\dagger Y_{u2}' Y_{u2}'^\dagger Y_{u1}'+Y_{u2}'^\dagger Y_{u1}'Y_{u1}'^\dagger Y_{u2}'\right)+2\left(Y_{u1}'^\dagger Y_{d2}' Y_{d2}'^\dagger Y_{u1}'+Y_{u2}'^\dagger Y_{d1}'Y_{d1}'^\dagger Y_{u2}'\right)\bigg],
\end{aligned}
\end{equation}
\begin{equation}
\begin{aligned}
\delta_{12}m^2_{\bar d}&=\frac{\Lambda^2}{(4 \pi)^4}\bigg[2Y_{d1}'^\dagger Y_{d2}'\Tr\left(3Y_{d2}'^\dagger Y_{d1}'+Y_{e2}'^\dagger Y_{e1}'\right)+2Y_{d2}'^\dagger Y_{d1}' \Tr\left(3Y_{d1}'^\dagger Y_{d2}' +Y_{e1}'^\dagger Y_{e2}' \right)\\
&\ \ \ \ \ +4\left(Y_{d1}'^\dagger Y_{d1}'Y_{d2}'^\dagger Y_{d2}'+Y_{d2}'^\dagger Y_{d2}' Y_{d1}'^\dagger Y_{d1}'\right)+2\left(Y_{d1}'^\dagger Y_{d2}' Y_{d2}'^\dagger Y_{d1}'+Y_{d2}'^\dagger Y_{d1}'Y_{d1}'^\dagger Y_{d2}'\right)\\
& \ \ \ \ \ +2\left(Y_{d1}'^\dagger Y_{u2}' Y_{u2}'^\dagger Y_{d1}'+Y_{d2}'^\dagger Y_{u1}'Y_{u1}'^\dagger Y_{d2}'\right)\bigg],
\end{aligned}
\end{equation}
\begin{equation}
\begin{aligned}
\delta_{12}m^2_{L}&=\frac{\Lambda^2}{(4 \pi)^4}\bigg[Y_{e1}'Y_{e2}'^\dagger \Tr\left(3Y_{d2}' Y_{d1}'^\dagger +Y_{e2}'Y_{e1}'^\dagger \right)+Y_{e2}'Y_{e1}'^\dagger \Tr\left(3Y_{d1}' Y_{d2}'^\dagger +Y_{e1}' Y_{e2}'^\dagger \right)\\
&\ \ \ \ \ +2\left(Y_{e1}' Y_{e1}'^\dagger Y_{e2}' Y_{e2}'^\dagger+Y_{e2}'Y_{e2}'^\dagger Y_{e1}' Y_{e1}'^\dagger\right)+\left(Y_{e1}' Y_{e2}'^\dagger Y_{e2}' Y_{e1}'^\dagger+Y_{e2}' Y_{e1}'^\dagger Y_{e1}'Y_{e2}'^\dagger\right)\bigg], \label{eq:correction2HiggsL}
\end{aligned}
\end{equation}
\begin{equation}
\begin{aligned}
\delta_{12}m^2_{\bar e}&=\frac{\Lambda^2}{(4 \pi)^4}\bigg[2Y_{e1}'^\dagger Y_{e2}'\Tr\left(3Y_{d2}'^\dagger Y_{d1}'+Y_{e2}'^\dagger Y_{e1}'\right)+2Y_{e2}'^\dagger Y_{e1}' \Tr\left(3Y_{d1}'^\dagger Y_{d2}' +Y_{e1}'^\dagger Y_{e2}' \right)\\
&\ \ \ \ \ +4\left(Y_{e1}'^\dagger Y_{e1}'Y_{e2}'^\dagger Y_{e2}'+Y_{e2}'^\dagger Y_{e2}' Y_{e1}'^\dagger Y_{e1}'\right)+2\left(Y_{e1}'^\dagger Y_{e2}' Y_{e2}'^\dagger Y_{e1}'+Y_{e2}'^\dagger Y_{e1}'Y_{e1}'^\dagger Y_{e2}'\right)\bigg].
\end{aligned}
\end{equation}
\endgroup
The MSSM-like Higgs will not get a correction from having multiple messengers outside of copies of the single messenger result. The $\tilde{A}$ terms are also unmodified. 

In the limit in which the Yukawas are all diagonal and real, and have only nonzero third family entries,
these corrections take the form
\begin{equation}
\begin{aligned}
\delta_{12}m^2_{Q}&=\frac{\Lambda^2}{128\pi^4}(6 Y_t'^2Y_t''^2+6 Y_b'^2Y_b''^2+Y_b''Y_b'Y_e'Y_e''+ Y_b'^2Y_t''^2+ Y_t'^2Y_b''^2),
\\
\delta_{12}m^2_{\bar u}&=\frac{\Lambda^2}{128\pi^4}(12 Y_t'^2Y_t''^2+Y_b'^2Y_t''^2+ Y_t'^2Y_b''^2),
\\
\delta_{12}m^2_{\bar d}&=\frac{\Lambda^2}{128\pi^4}(12 Y_b'^2Y_b''^2+2Y_b''Y_b'Y_e'Y_e''+Y_b'^2Y_t''^2+Y_t'^2Y_b''^2),
\\
\delta_{12}m^2_{L}&=\frac{\Lambda^2}{128\pi^4}( 4 Y_\tau'^2Y_\tau''^2+3 Y_\tau''Y_\tau'Y_b''Y_b'),\\
\delta_{12}m^2_{\bar e}&=\frac{\Lambda^2}{64\pi^4}(4 Y_\tau'^2Y_\tau''^2+3 Y_\tau''Y_\tau'Y_b''Y_b'),
\end{aligned}
\end{equation}
in which we have used $Y_{t,b,\tau}'$ to denote the nonzero entries of $Y_{u1,d1,e1}$, and similarly $Y_{t,b,\tau}''$ for the nonvanishing entries of  $Y_{u2,d2,e2}$.

As a trivial example of the consistency of these results, let us consider a simplified scenario with only leptons and messengers coupling in the superpotential, with messenger Yukawas that are degenerate and diagonal.  The effective superpotential is
\begin{align}
W&=Y_eL\bar eH_d+Y_{e}'L\bar e M_{d1}+Y_{e}'L\bar e M_{d2}+(M+\theta^2F)M_{ui}M_{di}.
\end{align}
We introduce two new linear combinations of messengers 
\begin{align}
\Phi_{u/d}&=\frac 1 {\sqrt 2}\left(M_{u/d1}+M_{u/d2}\right),&\Theta_{u/d}&=\frac 1 {\sqrt 2}\left(-M_{u/d1}+M_{u/d2}\right),
\end{align} 
such that
\begin{align}
W&=Y_eL\bar eH_d+\sqrt{2}Y_{e}'L\bar e \Phi_{d}+(M+\theta^2F)\left(\Phi_{u}\Phi_d+\Theta_u\Theta_d\right).
\end{align}
The rotation has decoupled the $\Theta$ fields, leaving us with messenger-matter mixing through $\Phi$, resulting in the single messenger case with an additional factor of $\sqrt2$. We can look at the structure of the single messenger corrections $\delta m_{\tilde{L}}^2$ for instance, and notice that the terms that are quadratic in the messenger Yukawas will not contribute to the correction from two sets of messengers, but terms quartic in messenger couplings will have nontrivial contributions. These quartic contributions are
\begin{align}
\delta m_{\tilde{L}}^2(\sqrt2Y')&\supset \frac{\Lambda^2}{(4\pi)^4} \left[3 (\sqrt2 Y_e')^4+\Tr\left(3(\sqrt{2}Y_d')^2+(\sqrt2Y_e')^2\right)(\sqrt{2}Y_e')^2\right],\\
&\rightarrow \frac{\Lambda^2}{(4\pi)^4} \left[6Y_e'^4+2\Tr\left(3Y_d'^2+Y_e'^2\right)Y_e'^2\right]
+\frac{\Lambda^2}{(4\pi)^4} \left[6Y_e'^4+2\Tr\left(3Y_d'^2+Y_e'^2\right)Y_e'^2\right].\nonumber
\end{align}
The second term agrees with \eqref{eq:correction2HiggsL} taken in the same limit.

\bibliographystyle{prsty}

\end{document}